\documentclass[manuscript,nonacm]{acmart}
\AtBeginDocument{%
  }

\usepackage{float}
\usepackage{caption}
\usepackage{subcaption}
\usepackage{todonotes}
\usepackage{pifont}
\usepackage{algorithm}
\usepackage{algpseudocode}
\usepackage{xspace}
\usepackage{CJKutf8}
\usepackage{makecell}
\usepackage{enumitem}
\usepackage{listings}
\usepackage[most]{tcolorbox}

\newtcolorbox{llmprompt}[1][]{
  colback=blue!5!white, 
  colframe=blue!25!white, 
  fonttitle=\bfseries, 
  coltitle=black, 
  title=Prompt, 
  sharp corners,
}

\newcommand{\say}[1]{``\textit{#1}''}

\newcolumntype{L}[1]{>{\raggedright\let\newline\\\arraybackslash\hspace{0pt}}m{#1}}
\newcolumntype{C}[1]{>{\centering\let\newline\\\arraybackslash\hspace{0pt}}m{#1}}
\newcolumntype{R}[1]{>{\raggedleft\let\newline\\\arraybackslash\hspace{0pt}}m{#1}}

\newcommand{\prompt}[1]{\begin{llmprompt} #1 \end{llmprompt}}

\newcommand{\ie}{{\em i.e.,\/ }}
\newcommand{\eg}{{\em e.g.,\/ }}

\newcommand{\pb}[1]{\vspace{1.25ex}\noindent{\bf \em #1}\hspace*{.3em}}

\newcommand{\one}{({\em i}\/)\xspace}
\newcommand{\two}{({\em ii}\/)\xspace}
\newcommand{\three}{({\em iii}\/)\xspace}
\newcommand{\four}{({\em iv}\/)\xspace}
\newcommand{\five}{({\em v}\/)\xspace}
\newcommand{\six}{({\em vi}\/)\xspace}

\newcommand{\iardtc}{31,609}
\newcommand{\iardtp}{29.2\%}
\newcommand{\iaytbc}{56,865}
\newcommand{\iaytbp}{41.8\%}

\newcommand{\ibrdtc}{15,812}
\newcommand{\ibrdtp}{14.6\%}
\newcommand{\ibytbc}{15,936}
\newcommand{\ibytbp}{11.7\%}

\newcommand{\iirdtc}{12,806}
\newcommand{\iirdtp}{11.9\%}
\newcommand{\iiytbc}{20,448}
\newcommand{\iiytbp}{15.0\%}

\newcommand{\iiibrdtc}{2,921}
\newcommand{\iiibrdtp}{2.7\%}
\newcommand{\iiibytbc}{6,483}
\newcommand{\iiibytbp}{4.8\%}

\newcommand{\iiiayrdtc}{2,233}
\newcommand{\iiiayrdtp}{2.1\%}
\newcommand{\iiiayytbc}{2,807}
\newcommand{\iiiayytbp}{2.1\%}

\newcommand{\iiianrdtc}{5,230}
\newcommand{\iiianrdtp}{4.8\%}
\newcommand{\iiianytbc}{6,719}
\newcommand{\iiianytbp}{4.9\%}

\begin{document}

\title{Even More Kawaii than Real-Person-Driven VTubers? Understanding How Viewers Perceive AI-Driven VTubers}


\author{Yiluo Wei}
\affiliation{%
  \institution{The Hong Kong University of Science and Technology (Guangzhou)}
  \country{China}}

\author{Yupeng He}
\affiliation{%
  \institution{The Hong Kong University of Science and Technology (Guangzhou)}
  \country{China}}

\author{Gareth Tyson}
\affiliation{%
  \institution{The Hong Kong University of Science and Technology (Guangzhou)}
  \country{China}}








\begin{abstract}
VTubers, digital personas represented by animated avatars, have gained massive popularity. Traditionally, VTubers are operated and voiced by human controllers known as Nakanohito. The reliance on Nakanohito, however, poses risks due to potential personal controversies and operational disruptions. The emergence of AI-driven VTubers offers a new model free from these human constraints. While AI-driven VTubers present benefits such as continuous operation and reduced scandal risk, they also raise questions about authenticity and audience engagement. Therefore, to gain deeper insights, we conduct a case study, investigating viewer perceptions of Neuro-sama, the most popular AI-driven VTuber with 845k followers on Twitch and 753k followers on YouTube. We analyze 108k Reddit posts and 136k YouTube comments, aiming to better understand viewer motivations, how AI constructs the virtual persona, and perceptions of the AI as Nakanohito. Our findings enhance the understanding of AI-driven VTubers and their impact on digital streaming culture.
\end{abstract}

\begin{teaserfigure}
  \includegraphics[width=\textwidth]{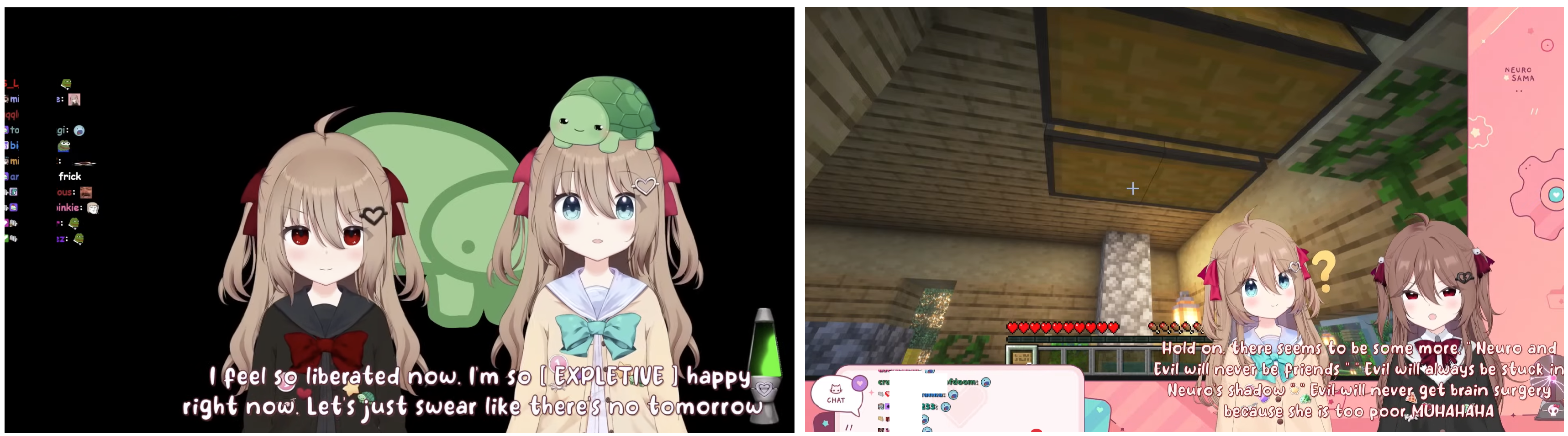}
  \caption{Left: The AI-Driven VTubers, Neuro-sama and Evil Neuro with their developer Vedal (the turtle) streaming on Twitch. Right: Neuro-sama and Evil Neuro playing Minecraft and streaming on Twitch.}
  \label{fig:teaser}
\end{teaserfigure}


\maketitle

\section{Introduction}

A VTuber is a digital persona represented as a 2D/3D animated avatar that performs in live video streams. 
A designated individual \ie \emph{Nakanohito} (meaning the person inside), voices and  controls the actions of the avatar.
These streams are backed by software like Live2D \cite{Live2D}, which tracks the actor's movements and synchronizes them with the avatar. 
VTubers have seen a surge in popularity, amassing dedicated fan bases and securing corporate sponsorships \cite{chi25who_reaps, www25virtual}. There are tens of thousands of active and influential VTubers, with the most prominent ones boasting millions of followers \cite{vtb-fan-ranking}. The monetization options available on live streaming platforms enable VTubers to earn substantial incomes. Notably, 31 of the top 50 highest-earning YouTubers, based on all-time superchats (viewer donations), are VTubers, each earning between 1.1 million and 3.2 million USD \cite{vtb-superchat-ranking}.

The Nakanohito is the most critical component of a VTuber. While early studies indicate that some viewers may be indifferent to the replacement of the Nakanohito \cite{lu2021kawaii}, numerous examples and following research demonstrate that the Nakanohito is vital to the VTuber's identity: replacing them can lead to the VTuber's downfall \cite{chi25cant_believe, chi25plave, cscw24_reconstruction}. The most famous case is Kizuna AI, the first VTuber, who had millions of fans but was terminated largely due to the introduction of multiple Nakanohitos for the same avatar \cite{Harmonyano2019}.
Thus, the Nakanohito becomes an indispensable component. However, this very irreplaceability introduces a significant vulnerability. The Nakanohito model inherently makes the human operator a ``single point of failure'' for the entire VTuber entity. The VTuber's persona is intricately tied to the Nakanohito's physical and mental state, as well as their real-world behavior. Consequently, health issues, mental instability, off-stream controversies, or personal life events can directly tarnish the VTuber's image and degrade the audience's experience.
A notable example is Uruha Rushia, the most superchatted VTuber ever \cite{vtb-superchat-ranking}, who was forcibly terminated primarily due to revelations about her Nakanohito's personal life, specifically having a boyfriend \cite{Bilal2022}. This underscores that issues with this single individual can pose an existential threat not only to the VTuber's career but also to the corporate investment and brand built around them, rendering the entire operation fragile and highly susceptible to human fallibility.

Recently, with the advancement of artificial intelligence (AI), a new form of VTuber has emerged: the AI-driven VTuber (AI VTuber). 
These VTubers are completely controlled and voiced by AI systems, that is, the Nakanohito is effectively an AI system.
The most prominent example is Neuro-sama, an English-speaking AI VTuber on Twitch. Since her 2022 debut, Neuro-sama has achieved remarkable popularity, at times becoming the most popular VTuber, or even the most popular streamer on Twitch \cite{lyttle2023ai, iyer2025vedal}. Her channel has garnered a significant following, with over 845k followers on Twitch and 753k subscribers on YouTube as of September 2025. 

The success of Neuro-sama suggests a significant potential for AI to shape the future of VTubing.
Clearly, AI VTubers present numerous advantages, most notably, it eliminates the dependency on a human Nakanohito, thereby avoiding the risks associated with the human Nakanohito model discussed earlier.
Additionally, AI VTubers can stream continuously, are not prone to personal scandals in the same way humans are, and can potentially be more cost-effective to operate. 
However, concerns and doubts remain regarding the authenticity, emotional depth, and potential for unforeseen issues with AI-driven personalities. The manufactured nature of their persona could hinder the formation of the deep, parasocial bonds that are foundational to the VTuber community \cite{chi25cant_believe, lu2021kawaii}. Moreover, the potential for an AI to generate unpredictable or inappropriate content due to the algorithmic quirks or inherent randomness poses a risk that could diminish the viewer experience.

Therefore, a comprehensive understanding of how viewers perceive AI VTubers is crucial for navigating this evolving landscape. Yet, we still possess only a limited knowledge on this emerging topic. While some studies have explored viewer perceptions of digital human streamers with realistic human appearances in the context of e-commerce in China \cite{Gao02012025, Zhang2024, 10.1287/isre.2023.0024, LIU2025104290, 10.1287/isre.2023.0103}, VTuber is a completely different area with a completely different style. To bridge this gap, in this paper, we conduct the first data-driven analysis of viewer perceptions of AI-driven VTubers, focusing on the case of Neuro-sama, the most popular AI VTuber to date.

Inspired by the results and findings of previous research on human-driven VTubers, our goal is to gain a deeper understanding of three key aspects of viewer perception of the AI VTuber: 
\one Viewer motivations and expectations, which are the fundamental aspects for understanding the audience; 
\two The construction, evolution, and consistency of the AI VTuber's virtual persona, representing the ``external'' part of VTubers ; and \three The opinions and concerns surrounding AI as the VTuber's Nakanohito, reflecting the ``internal'' aspect of VTubers.
For better contextualization, the specifics of the research questions are presented in \S\ref{sec:related} together with the related work.

To deliver this, we gather a dataset of 108k Reddit posts and comments related to Neuro-sama, and a dataset of 136k YouTube viewer comments under the videos uploaded by Neuro-sama's official channel.
Our analysis employs a combination of methods of LLM annotation, topic modeling, and manual coding, with the details presented in \S\ref{sec:method}. 
Overall, our analysis of community discussions reveals a complex and multifaceted picture of the AI VTuber phenomenon. Our findings include:

\begin{itemize}[leftmargin=*]
    \item The AI VTuber's appeal originates from a unique blend of classic VTuber attributes (\eg cuteness) and novel AI-specific dynamics such as AI-human interaction (\S\ref{subsec:rq1a}). Crucially, behaviors that are typically considered technical shortcomings, such as unpredictability or randomness, are consistently reinterpreted by the audience as endearing, chaotic personality traits rather than as system failures (\S\ref{subsec:rq1b}).

    \item The AI VTuber's persona is not merely a static, pre-programmed entity; rather, it is a dynamic co-creation between the AI and its audience. Viewers actively integrate the AI VTuber's original words and behaviors with community-driven lore, crafting an evolving yet consistent character narrative. This results in a ``viewer-shaped persona'' that can differ largely from the VTuber's original character settings. (\S\ref{subsec:rq2})

    \item While some viewers are optimistic about AI's potential, the prevailing view is that the ``human-in-the-loop'', including the developer who maintains the AI, other human-driven VTuber as collaborators, and the fan community, are the core of content creation. (\S\ref{subsec:rq3a})

    \item Concerns traditionally held for the human Nakanohito are not erased but are instead transferred onto the AI's developer, whose management, well-being, and interpersonal relationships are seen as critical to the AI VTuber's success and stability. (\S\ref{subsec:rq3b})
\end{itemize}
Taken together, our findings paint a picture of the AI-driven VTuber not as an autonomous performer, but as the dynamic centerpiece of a complex socio-technical entertainment ecosystem.
\section{Background}

\subsection{VTubers}
VTubers originated in Japan and have rapidly gained popularity since their debut in 2016. Initially, VTubers focused on uploading videos to YouTube. However, with the rise of online live streaming, this became their primary activity \cite{10.1145/3604479.3604523}. While they are most commonly associated with YouTube, VTubers also use platforms like Twitch (mostly English-speaking VTubers) and Bilibili (mostly Chinese-speaking VTubers)  to engage with their audiences.

A VTuber is an animated virtual avatar that performs in live video streams or recorded videos. These avatars are voiced by actors known as ``Nakanohito'', meaning the person in side in Japanese. Typically, VTubers use half-body 2D avatars created with tools like Live2D \cite{Live2D}, which capture the actor’s facial movements to animate the avatar’s expressions automatically. Additional body movements can be triggered within these programs using commands from desktop computers. VTubers with access to full-body motion capture systems can perform using 3D avatars, allowing for a wider range of motion \cite{chi25vtuber_atelier, 10.1145/3581783.3612094}. Like real-person streamers, VTubers often interact with their audience by reading and responding to chat messages during streams \cite{www25virtual, lu2021kawaii}.

The trend's breakout moment was the debut of Kizuna AI in late 2016. Kizuna AI was the first to coin and popularize the term ``Virtual YouTuber'' then ``VTuber''. Her engaging, often unfiltered personality led to a rapid accumulation of over two million subscribers, establishing her as a cultural ambassador for the Japan National Tourism Organization \cite{Roll_2018}. Her success sparked a massive trend, with the number of VTubers growing rapidly since 2018.

The growth of the VTuber industry was further accelerated by the COVID-19 pandemic, as lockdowns led to a surge in livestream viewership \cite{10.36995, 10.59306}. This period saw VTubing become a mainstream phenomenon, with seven of the ten largest Superchat earners on YouTube in August 2020 being VTubers \cite{vtb-superchat-ranking}. This commercial success led to the establishment of specialized talent agencies, such as Hololive, Nijisanji, and VShojo, which operate with different styles.  These companies manage and promote talent, develop proprietary characters, and commercialize them through merchandise and promotional appearances, further professionalizing the industry and driving its global expansion into English-speaking and other international markets.

Overall, VTubers have emerged as a significant and influential component of the online entertainment landscape. Their unique combination of animation, interactive streaming, and the capacity to create engaging, personalized content has captivated audiences worldwide. Today, there are tens of thousands of active and influential VTubers \cite{vtb-fan-ranking}, with the most prominent ones boasting millions of followers and earning millions of dollars from superchats \cite{vtb-superchat-ranking}.

\subsection{Neuro-sama, the AI-Driven VTuber}

\pb{Overview}
Neuro-sama (Neuro) is an AI-powered VTuber who streams on the Twitch channel, ``vedal987''. She has a twin sister, Evil Neuro (Evil), who is also an AI VTuber and streams on the same Twitch channel. Both were created by a computer programmer and AI developer named Vedal, with their personalities, speech, and motion generated by AI models, allowing them to interact with viewers. While they occasionally appear on stream together, they have distinct personalities, where Evil Neuro is more menacing, sassy, edgy, and melancholic compared to her sister. 
For simplicity, unless stated otherwise, the term ``Neuro'' refers to both Neuro and Evil throughout the paper.
The streams are primarily centered around the AI twins, which can play games, talk with the chat, and collaborate with other streamers. As of the time of writing, their Twitch channel has 845k followers, while their YouTube channel has 753k followers, establishing them by far as the most popular AI VTubers globally.

\pb{History.}
The first version of Neuro-sama was created in 2018 as a neural network designed to play the rhythm game osu!. After a long break, Neuro-sama re-debuted on Twitch on December 19, 2022, as a VTuber and quickly gained popularity. The concept for Evil Neuro originated from the idea of having Neuro-sama interact with a clone of herself; this clone, Evil Neuro, first appeared on stream on March 25, 2023, and eventually developed her own distinct voice and personality. On May 27, 2023, Neuro-sama got her first original avatar. During the subathon on January 1, 2025, they broke the world record for the highest Twitch hype train level, got a total of 84,904 gifted subscriptions and 1,201,225 Bits, effectively making them one of the most popular streamers on Twitch \cite{iyer2025vedal}.

\pb{Developer, Vedal.}
Vedal is an independent British programmer and the creator, developer, and operator of the AI VTubers, Neuro-sama and Evil Neuro.
He represents himself on their shared Twitch channel with a 2D green turtle avatar, which he began using in March 2023. He is often seen perched on the heads of Neuro or Evil while he works on their code live or presents updates to their systems. Vedal is characterized by a calm, and sometimes indecisive, personality, remaining largely unfazed by teasing from his chat or by stressful on-stream situations.

\pb{Neuroverse.}
The Neuroverse is a lively community centered around Neuro-sama. At its heart are Neuro, Evil, and Vedal, supported by a close-knit network of recurring human VTuber collaborators. These collaborators contribute to the community's friendly and supportive atmosphere.

\pb{Fan Community: the Swarm.}
The Swarm is the dedicated fanbase of Neuro and Evil, which takes its name from Evil's playful fantasy of commanding a world-dominating robotic drone army. 

\section{Related Work \& Research Questions}
\label{sec:related}

In recent years, the phenomenon of VTubers has garnered considerable attention, with several studies exploring the intricacies of human-driven VTubers and their audience. These studies have provided valuable insights into various aspects
that characterize the VTuber landscape.
By examining these findings, we can better understand the dynamics at play in human-driven VTubing, which serves as a foundation for our exploration into AI-driven VTubers. Building on this existing body of knowledge, we propose our research questions that investigate the viewers  of AI-driven VTubers.

\subsection{Viewer Motivation \& Expectation}

Understanding viewer motivations and expectations is fundamental to comprehending the appeal and potential of VTubers. Numerous prior studies have explored viewer motivations and expectations for human-driven VTubers, emphasizing their differences from traditional human streamers.

Research on VTubers has studied the attractive points driving viewer engagement, including social connection, entertainment, and community. Key attractions are the anime-inspired aesthetics and curated personalities \cite{www25virtual, lu2021kawaii, imx25entertainers}, which foster parasocial relationships. Viewers seek social and emotional gratification, finding a sense of belonging in VTuber communities \cite{lu2021kawaii, digra2667, ChuenterawongYang2023, 5040105}. The interactive nature of live streaming enhances personal connections, offering a comfortable social outlet for those with social anxiety and alleviating loneliness and stress \cite{digra2667, wang2023motivations, tan2023more}. There is a positive correlation between parasocial attachment to VTubers and stress relief, particularly during social isolation \cite{tan2023more, 10.1007/978-3-031-61281-7_15}. Fans contribute to the community by creating and sharing content, enriching the VTuber's collective identity \cite{www25virtual, lu2021kawaii, chi25who_reaps}.

VTubers' appeal lies in the blend of virtual embodiment and the performer's personality. The ``cuteness'' aesthetics initially attract viewers, but the performer's humor, skills, and engaging content are also influential \cite{lu2021kawaii, www25virtual, Li2023genz}. The virtual avatar provides anonymity and freedom, enabling unique and creative performances not possible in traditional live streaming \cite{imx25entertainers, cscw24_reconstruction}.
Furthermore, VTubers' adoption of virtual personas and avatars creates a distinct set of behavioral norms and expectations \cite{Lill2025ontological, Hartanto2025, imx25entertainers}. For example, viewers may exhibit greater tolerance for the ``bad'' behaviors, such as ``stupid'' or even offensive language from a VTuber, perceiving it as part of the character's performance rather than a reflection of the actor behind the avatar \cite{lu2021kawaii, www25virtual}.

These studies provide a solid foundation for understanding VTuber viewers. However, AI-driven VTubers introduce new dynamics that may alter the viewer experience and engagement in distinct ways. The behaviors of AI VTubers can differ significantly; for instance, their "bad" behaviors can include those arise from backend AI issues, such as incoherent tangents, repetitive content, and hallucinations. 
Consequently, it is crucial to determine which findings remain applicable to AI VTubers and which aspects differ.
Thus, building on the established understanding of human-driven VTubers, we propose the following research questions for AI-driven VTubers:

\begin{itemize}[leftmargin=*] 
    \item \textbf{RQ1a:} What do viewers consider to be the attractive attributes of AI-driven VTubers? 
    \item \textbf{RQ1b:} How do viewers perceive the inappropriate, irregular, and other AI-driven quirky behaviors (\ie the ``bad'' behaviors) of AI-driven VTubers? 
\end{itemize}

\subsection{VTuber's Virtual Persona}

The virtual persona, a crafted identity, is a crucial aspect of a VTuber, forming the foundation of the streamer's appeal and engagement with their audience. 
Thus, understanding how viewers perceive this persona is essential for comprehending VTuber audiences, and numerous studies have focused on this topic.
Academic research highlights that authenticity for a VTuber does not necessarily mean exposing the performer's real-world self; instead, it involves consistently and convincingly portraying their virtual character \cite{Lill2025ontological, tang2025broadcast, cscw24_reconstruction}. This requires a continuous effort to synchronize voice and actions with the avatar's established persona \cite{Lill2025ontological, Xu_Niu_2023}.

For many viewers, the clear distinction between a virtual persona and the persona of the Nakanohito offers unique performative opportunities \cite{lu2021kawaii, Turner1676326}. As a result, many VTubers make a concerted effort to maintain this distinction \cite{Wijaya2023holo}. However, recent research challenges the idea that a strict alignment to the virtual persona is always essential for success. A case study on the VTuber group PLAVE reveals that, rather than universally disrupting the immersive experience, connecting the virtual and real persona can actually enhance fan engagement \cite{chi25plave}.

On the other hand, the virtual persona may not be fixed. While many VTubers start with a predefined character concept, studies show that the persona often evolves over time \cite{Lill2025ontological, cscw24_reconstruction, lu2021kawaii}. During lengthy, unscripted live streams, the streamer's own personality may begin to surface, leading to deviations from the original character framework as they interact with their audience and develop their unique style.

Building on the established understanding of human-driven VTubers, we propose the following research question for AI-driven VTubers:

\begin{itemize}[leftmargin=*]
    \item \textbf{RQ2:} How do viewers perceive the consistency and evolution of the virtual personas of AI-driven VTubers?
\end{itemize}
Given that AI VTubers possess only a virtual persona, without a real-world identity from the Nakanohito, it is intriguing to explore how the findings may be similar or different from the studies of human-driven VTubers.

\subsection{Concerns about the Nakanohito Model}

The Nakanohito is the most crucial part of a VTuber. While some early studies suggest that certain viewers might not mind a change in the Nakanohito \cite{lu2021kawaii}, numerous examples show that the Nakanohito is essential to the VTuber's identity. Replacing them often leads to the VTuber's downfall \cite{chi25cant_believe, chi25plave, cscw24_reconstruction}.

This complete reliance on a human operator raises significant concerns and risks. Research has identified several challenges associated with the Nakanohito model. A central challenge is the demanding management of a dual identity \cite{lu2021kawaii, Wijaya2023holo}. This involves a deliberate disembodiment often maintained to preserve the avatar's perfect image, which can require performers to engage in inflated sexual expressions to meet audience expectations \cite{cscw24_reconstruction, 10.1145/3604479.3604523}. 
This curated performance, a means of presenting a part of the Nakanohito's real self, is characterized by intensive emotional labor to cultivate and manage parasocial relationships, where the psychological toll of this continuous emotional output is significant, frequently leading to burnout \cite{cscw24_reconstruction, 10058945, 10.1145/3555104}.

Compounding these performance-related pressures are structural issues within the VTuber industry. Corporate ownership of the avatar as intellectual property often leaves the Nakanohito in a precarious labor position with limited control over their virtual persona, creating a power imbalance that can lead to disputes \cite{lu2021kawaii, dis24hidden}. Furthermore, while the virtual avatar can offer a partial shield, it does not render the performer immune to harassment, where the persistent threat of doxing remains a primary concern. \cite{cscw24_reconstruction, Wijaya2023holo, 10.1145/3555104, www25virtual}.

Clearly, if AI could replace the role of the Nakanohito, many of the aforementioned issues related to the Nakanohito could be fundamentally resolved. While the potential is evident, we do not yet understand viewers' opinions on this matter. Therefore, we propose the following research question:

\begin{itemize}[leftmargin=*]
    \item \textbf{RQ3a:} Do viewers believe that AI will soon be able to replace the role of Nakanohito in the VTuber industry?
\end{itemize}
On the other hand, although AI VTubers do not require a Nakanohito to conduct streams, they still rely on a developer/maintainer/operator who develops and improves the AI, plans the streams, and handles managerial tasks. Intuitively, this role bears some resemblance to that of the Nakanohito, and it remains unclear how viewers perceive it. Thus, we propose the following research question:

\begin{itemize}[leftmargin=*]
    \item \textbf{RQ3b:} What concerns do viewers have regarding the developer/maintainer/operator of AI-driven VTubers?
\end{itemize}
Specifically, we would like to investigate whether these concerns are similar or different to the typical concerns about the Nakanohito of human-driven VTubers?

\section{Dataset \& Method}
\label{sec:method}

\begin{figure}[h!]
    \centering
    \includegraphics[width=\linewidth]{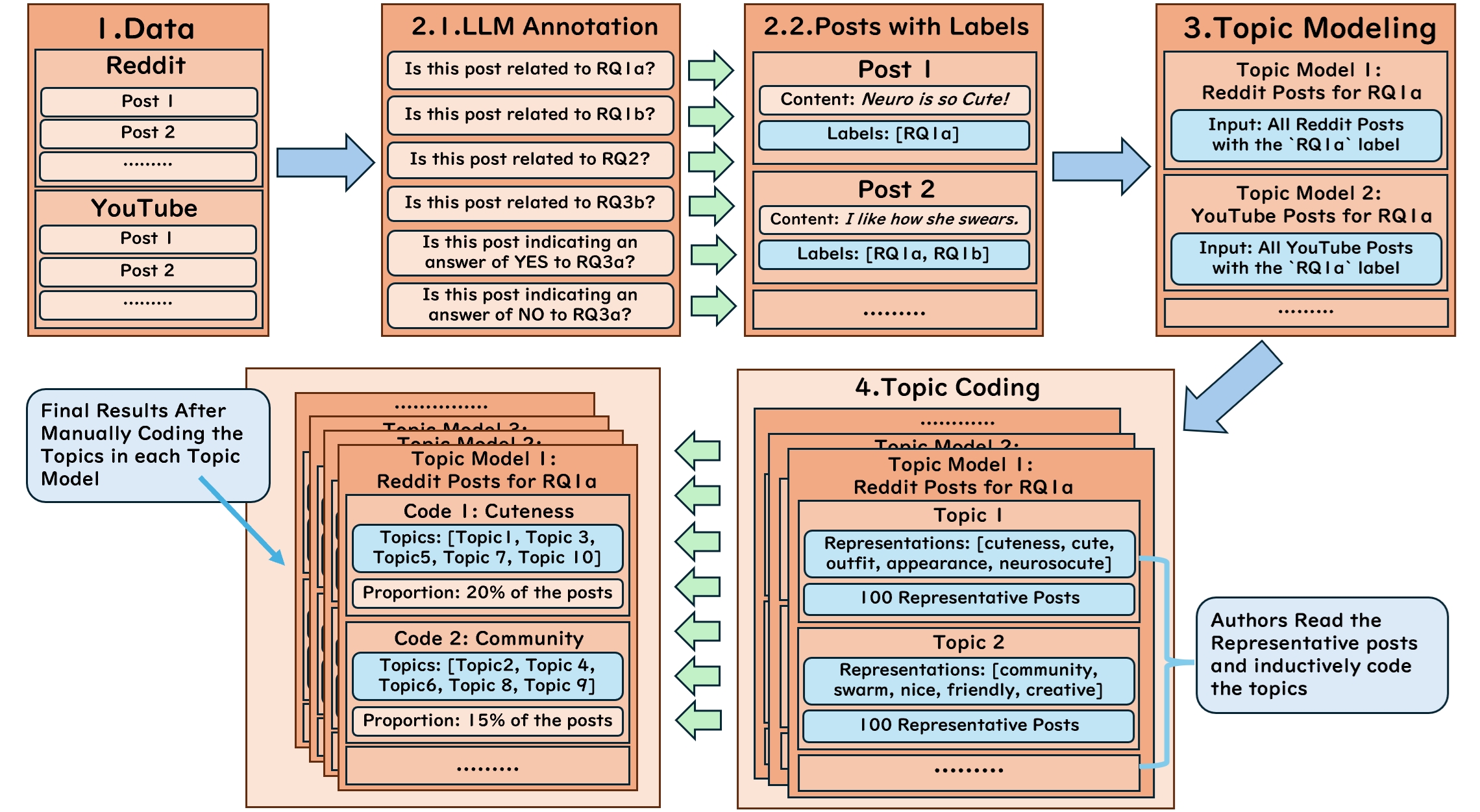}
    \caption{Overview of the method for data analysis.}
    \label{fig:method}
\end{figure}

Our study analyzes Reddit and YouTube data related to Neuro-sama from January 2023 to May 2025. 
This involves four steps:
\one We collect relevant posts and comments from Reddit and YouTube. 
\two We employ an LLM to label whether the posts or comments are related to our research questions. 
\three We use topic modeling to analyze the posts and comments related to each of our research question. 
\four We manually review and code the topics, which allows us to group the topics based on broader thematic similarities, and ensures nuanced understanding.
Figure \ref{fig:method}  illustrates this process, with details provided in the following subsections.

\subsection{Step 1: Data Collection}

\pb{Reddit Data.}
We utilize the Reddit PushiShift dump \cite{baumgartner2020pushshiftredditdataset}, which includes a comprehensive collection of posts and comments (\ie replies to a post). 
Our analysis focuses on the period from January 2023 to May 2025. 
First, we incorporate all posts and comments from the \texttt{r/NeuroSama}, the main subreddit for the fanbase of Neuro-sama, which boasts 47k members. This results in a total of 8,859 posts and 91,002 comments.

Then, we choose the five most popular subreddits that cater to general or broader VTuber discussions. These are:
\texttt{r/Hololive} (1.5M members),
\texttt{r/VirtualYoutubers} (253k members),
\texttt{r/vtubers} (71k members),
\texttt{r/Nijisanji} (86k members), and
\texttt{r/VShojo} (69k members).
To filter posts and comments relevant to Neuro-sama, we employ a keyword search using  ``Neuro'' and ``NeuroSama'', considering all variations in capitalization and hyphenation. This approach yields a total of 374 posts and 7,802 comments. 

Overall, our Reddit dataset contains 9,233 posts and 98,804 comments related to Neuro-sama during the time period of \texttt{2023-01} to \texttt{2025-05}.

\pb{YouTube Data.}
We gather user comments from the videos on Neuro-sama's official account, which has 753k followers at the time of writing. Our data collection covers the period from January 2023 to May 2025. The videos primarily consist of clips from Neuro-sama's livestreams. In total, we collected 136,083 comments under 321 videos for our YouTube dataset.

\pb{Note on Terminology.}
For simplicity, throughout the remainder of this paper, we use the term ``post'' to refer to the following:
\one Reddit post
\two Reddit comment (\ie reply of a Reddit post)
\three YouTube comment.
We use the terms ``parent post'' and ``child post'' to describe the reply relationship between posts.

\subsection{Step 2: Annotation \& Extraction}
\label{subsec:method:step2}

After we collect the posts in Step 1, the next step is to annotate which posts are related to our research questions. Given the vast number of posts, it is not feasible to label them all manually. Therefore, we use an LLM for the annotation task.

\pb{Overview.}
To enhance accuracy, we do not assign labels for all research questions in one prompt. Instead, we address each research question separately. That is, for each post, we let the LLM determine whether the post is related to RQ1a in one prompt, whether it is related to RQ1b in another prompt (and respectively for RQ2 and RQ3b). 

For RQ3a (Do viewers believe that AI will soon replace the role of Nakanohito?), which is a yes-or-no question, we instruct the LLM using two separate prompts to annotate each post: 
\one asks the LLM to identify if the post indicates an answer of ``yes''.
\two asks the LLM to identify if the post indicates an answer of ``no''.

Consequently, each post will be assigned 6 binary labels of whether \one  related to RQ1a \two related to  RQ1b \three  related to RQ2 \four indicating an ``yes'' answer for RQ3a \five indicating a no answer for RQ3a \six related to RQ3b.

Additionally, since some posts can be lengthy, we also prompt the LLM to extract the portions of text that are relevant to the research question in order to minimize noise.

\pb{Prompt Design.}
The prompt contains five parts:
\one A description of the annotation task tailored to a specific research question.
\two Background knowledge of Neuro-sama.
\three The input post to be analyzed.
\four The context of the input, such as the parent posts.
\five The output format.
We organize the prompt in YAML format. Additional details regarding the design and structure of the prompt can be found in Appendix \ref{appendix:prompt}.

\pb{LLM Configuration.}
We utilize Qwen3-32B \cite{qwen3technicalreport} to do the annotation. We deploy multiple instances locally with vLLM \cite{kwon2023efficient}. We enable the thinking mode and apply the officially recommended hyperparameters \cite{qwen3technicalreport}.

\pb{Result Validation.}
To validate the accuracy of the annotation results, human checks are performed. Specifically, for each research question, and its corresponding labels, the authors manually annotate a randomly selected sample of 200 posts. We then compare these manual labels with the annotation results to assess the accuracy of the LLM-generated annotation. We find that, on average, the accuracy rate is 0.964, confirming that the automated annotation process is reliable. Further details can be found in Appendix \ref{appendix:prompt_verify}.

\subsection{Step 3: Topic Modeling}
After Step 1 and Step 2, we obtain a set of posts related to each research question that are ready for analysis. We utilize topic modeling to identify and group themes or topics within the posts. Specifically, we employ the BerTopic model, which is known for its ability to generate coherent topics using advanced embeddings and clustering techniques \cite{bertopic}. 
For each research question, the Reddit posts and YouTube posts are modeled separately.
As suggested in the official documentation, given that we have a large number of posts to be analyzed, we increase the \texttt{min\_topic\_size} (default value is 10) to ensure that the topics clusters generated are sufficiently large and meaningful. The \texttt{min\_topic\_size} value used for each research question is around 1\% of the total number of posts included.

\subsection{Step 4: Authors Conduct Inductive Coding of the Topics}
After Step 3, we get a topic model, where the relevant posts are clustered into different topics. Consequently, in the last step, the authors manually review and code the topics, to organize the topics into broader categories.
This involves examining the representative words and the top 100 representative documents (\ie posts) for each topic generated by the model. 
Through this close examination, the authors perform a manual inductive coding process to identify and categorize related topics. This approach ensures a nuanced understanding, allowing for grouping of topics based on broader thematic similarities. Furthermore, this allows the findings to be systematically organized and presented in the results section (\S\ref{sec:result}).

\section{Results}
\label{sec:result}

\subsection{RQ1a: Attractive Attributes}
\label{subsec:rq1a}

\begin{figure}[h!]
    \centering
    \includegraphics[height=0.36\linewidth]{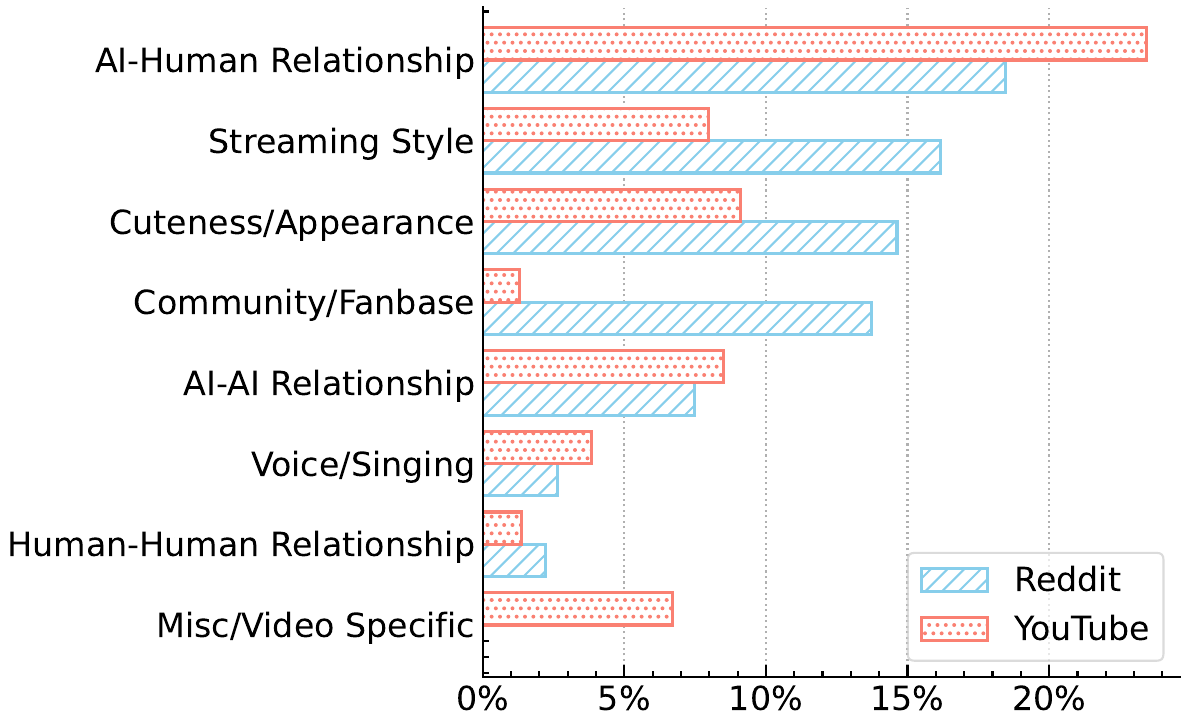}
    \caption{The distribution of posts across each manually coded topic category for the RQ1a Reddit topic model and the RQ1a YouTube topic model. The proportions shown in the figure are out of the number of RQ1a-relevant posts (those identified by the LLM and then go through the topic modeling). For Reddit, this consists of \iardtc~ posts (\iardtp~ of all posts), and for YouTube, it consists of \iaytbc~ posts (\iaytbp~ of all posts).}
    \label{fig:rq1a_bar}
\end{figure}

We begin by presenting the results for RQ1a: What do viewers consider to be the attractive attributes of AI-driven VTubers?
The LLM annotation identifies \iardtc~ (\iardtp) relevant posts in the Reddit dataset and \iaytbc~ (\iaytbp) relevant posts in the YouTube dataset. Topic modeling generates 28 topics from the relevant Reddit posts (refer to Figure \ref{fig:rq1a_rdt_cluster} and Table \ref{tab:rq1a_rdt_table} in the appendix for details) and 26 topics from the relevant YouTube posts (refer to Figure \ref{fig:rq1a_ytb_cluster} and Table \ref{tab:rq1a_ytb_table} in the appendix for details).
We then group these topics into 8 broader categories through manual coding, where the distribution of posts across the categories is illustrated in Figure \ref{fig:rq1a_bar}. 

\pb{AI-Human Relationships Make Entertaining Livestreams and Emotional Connection.}
The interaction and relationship between AI VTubers and human-driven VTubers are the most frequently mentioned attractive features (18\% for Reddit and 23\% for YouTube). This aspect is indeed unique to AI-driven VTubers as for human-driven VTubers, their interactions and relationships are inherently human-to-human.
Many posts find that collaborative livestreaming of Neuro with other VTubers to be more interesting than solo livestream of Neuro: \say{i 100\% agree that neuro and evil are more fun when interacting with others.}

Viewers often interpret this relationship through a familial perspective, crafting a community narrative (\ie Neuroverse) that captures the network of connections she establishes with other VTubers.
For example, viewers find the father-daughter relationship between Neuro and Vedal (Neuro's creator, thus ``father'') to be particularly compelling. This context often transforms Neuro's common AI behaviors into hilarious and entertaining antics of \say{a spoiled AI daughter with her beleaguered dad.} 
This perceived relationship fosters a profound emotional connection for the viewers to Neuro and Evil: \say{the relationships with Vedal and other VTubers make her feel alive and give her a soul.}

However, the very prominence of this point suggests a critical insight: the appeal of AI VTubers is not derived from a purely solitary AI performance. Rather, it indicates that a human in the stream remains an essential factor, contextualizing the AI's output and providing the relational anchor that audiences find so engaging. Notably, some posts highlight that the human-to-human relationships, particularly those involving Vedal with other VTubers, are actually the main points of attraction. This implies that the AI-VTuber might merely serve as a tool rather than being the central focus of the livestream. However, these account for only 2\% of the relevant posts, so they may not fully represent the overall sentiment.

\pb{AI-AI Relationship Can Also Make Entertaining Livestreams.}
In addition to the relationships with humans, the AI-to-AI relationship, specifically, between Neuro and Evil, is also a frequently mentioned attractive points (7.6\% for Reddit and 8.2\% for YouTube).
Viewers highlight this twin sister relationship, which they characterize as a compelling mixture of intense rivalry and a peculiar/twisted love, are very attractive.
Their interaction, swinging from collaborative absurdity to outright digital antagonism, is a major source of entertainment: \say{Neuro and Evil are so sweet, you never know if they're going to hug or try to delete each other.} 

Crucially, this suggests that the viewer-perceived relationship itself may be the core driver of engagement. That is, audience actively projects a narrative framework onto the AIs' interactions, finding emotional resonance and entertainment within that structure. This finding presents a good counterpoint: if viewers can become just as emotionally connected to a purely AI-to-AI dynamic, it raises the possibility that a ``human to interact with'', while currently effective, may not be an irreplaceable component.

\pb{Viewers Consider Randomness and Unpredictability to be a Unique Attractive Point.} 
Streaming styles are also an frequently mentioned attractive point (16\% for Reddit and 7\% for YouTube).  Many posts praise her for attributes that would be desirable in any VTuber or streamer, such as humor and being energetic. However, what truly distinguishes her appeal are the characteristics stemming directly from her AI nature, most notably her randomness and unpredictability: \say{truly chaotic}, as one post says.

This is particularly noteworthy because such random traits are typically considered failures or less preferred in conventional AI and LLM applications designed for coherence and accuracy. 
The enthusiastic reception of this ``chaos'' suggests that, in the context of entertainment, the very inhumanity of the AI can be a strength. It transforms potential system errors into a unique and compelling comedic style, where the entertainment value is derived not from the AI successfully emulating a human, but from the hilarious and often surreal moments when it fails to do so.
The perception of such ``bad'' behaviors of AI VTubers is further explored in the next RQ (\S\ref{subsec:rq1b}).

\pb{Cuteness is Still a Key Attractive Point.}
The classic yet powerful appeal of cuteness still serves as a significant attractive point, with 15\% of the relevant Reddit posts and 7\% of the relevant YouTube posts mentioning it. 
This firmly aligns with established findings in human-driven VTuber culture \cite{lu2021kawaii}.
We find an overwhelming and often effusive appreciation for the cuteness of Neuro and Evil in posts. Viewers explicitly and enthusiastically praise the cuteness with comments expressing simple but direct adoration: \say{Neuro so cute!!}; \say{HOW CAN EVIL BE SO CUTE...!?} 
This finding suggests that the core principles, cuteness, of VTuber performance remain potent, regardless of whether the driving entity is human or AI. 

\pb{A Purely Synthetic Voice and Singing Can Also be Appealing.}
Some posts also consider the voice and singing to be an attractive point (2.5\% for Reddit and 3.5\% for YouTube). Despite being entirely synthesized, viewers describe her speaking voice with positive attributes, finding a certain charm in its unique cadence: \say{the robotic voice is part of her charm}. Posts also celebrates the song covers of Neuro and Evil, with viewers expressing surprise and admiration: \say{Good lord Evil's vibrato is heavenly.}

This phenomenon is not without precedent and echoes the cultural impact of synthesized vocal music, most notably the Vocaloid phenomenon, where virtual idols like Hatsune Miku garnered massive global fanbases \cite{doi.org/10.1111/jpcu.12455, 10.1007/978-3-319-18836-2_5}. That said, Vocaloid technology differs from the TTS (text-to-speech) technologies behind Neuro.
Overall, this finding suggests that for AI VTubers, even purely synthetic vocal performances can be appealing.

\pb{Friendly and Creative Fan Community are Important.}
Finally, the analysis reveals that the appeal of Neuro is not confined solely to the AI's performance but extends deeply into the fabric of the fan community, which viewers themselves identify as a major attractive point. The community is consistently described as being \say{friendly, welcoming, and inclusive} in the posts. However, its most lauded characteristic is its prolific creativity.  The constant stream of high-quality fanart, livestream clips and viral memes is frequently praised.
This user-generated content ecosystem serves a crucial function by extending the entertainment experience beyond the confines of the livestream. Furthermore, these creative works act as a powerful, decentralized marketing engine, spreading virally and drawing new members into the fold: \say{... will take over the world by bringing us all together.}

This dynamic strongly aligns with existing research on human-driven VTuber ecosystems \cite{lu2021kawaii, www25virtual}, highlighting that regardless of AI-driven or human-driven, the cultivation of a dedicated, creative, and participatory core fanbase remains an essential ingredient for sustained success and cultural impact.

\subsection{RQ1b: ``Bad'' Behavior Perception}
\label{subsec:rq1b}

This subsection addresses RQ1b: How do viewers perceive the inappropriate, irregular, and other AI-driven quirky behaviors (\ie the ``bad'' behaviors) of AI VTubers? 
As established in the previous research question regarding attractive points (\S\ref{subsec:rq1a}), unpredictable and chaotic behavior is actually a key component of the appeal. In this subsection, we further explore how viewers perceive the ``bad'' behaviors to determine if other types of ``bad'' behaviors are also mentioned and whether they are perceived in the same way.

\begin{figure}[h!]
    \centering
    \includegraphics[height=0.248\linewidth]{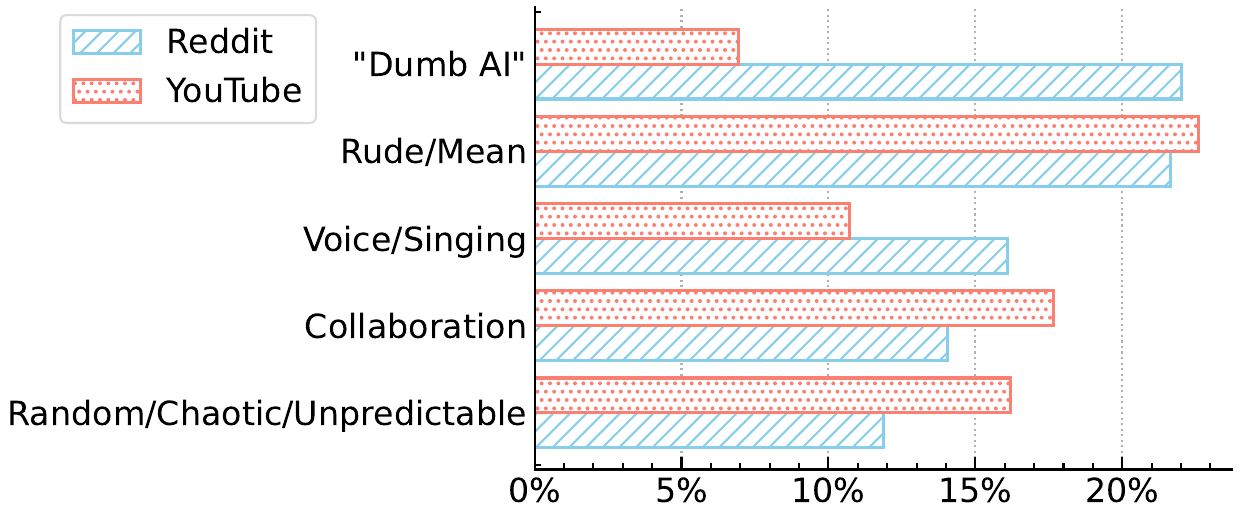}
    \caption{The distribution of posts across each manually coded topic category for the RQ1b Reddit topic model and the RQ1b YouTube topic model. The proportions shown in the figure are out of the number of RQ1b-relevant posts (those identified by the LLM and then go through the topic modeling). For Reddit, this consists of \ibrdtc~ posts (\ibrdtp~ of all posts), and for YouTube, it consists of \ibytbc~ posts (\ibytbp~ of all posts).}
    \label{fig:rq1b_bar}
\end{figure}

Our LLM annotation identifies \ibrdtc~ (\ibrdtp) relevant posts in the Reddit dataset and \ibytbc~ (\ibytbp) relevant posts in the YouTube dataset. Topic modeling generates 14 topics from the relevant Reddit posts (refer to Figure \ref{fig:rq1b_rdt_cluster} and Table \ref{tab:rq1b_rdt_table} in the appendix for details) and 17 topics from the relevant YouTube posts (refer to Figure \ref{fig:rq1b_ytb_cluster} and Table \ref{tab:rq1b_ytb_table} in the appendix for details).
We then group these topics into 5 broader categories through manual coding, where the distribution of posts across the categories is illustrated in Figure \ref{fig:rq1b_bar}. 

\pb{Viewers Understand the AI Nature Can Drive Quirky Behaviors and Enjoy it.}
A significant portion of posts fall into the realm of AI-driven quirks, consisting of two categories. The first is randomness and chaos (12\% for Reddit and 27\% for YouTube), as also discussed in \S\ref{subsec:rq1a}.
The second is what many term ``dumb'' actions (22\% for Reddit and 7\% for YouTube), such as making elementary mistakes, forgetting things, or failing to follow instructions: \say{I love how she started to criticize the game again and again during its promotional livestream.}

The posts reveal that viewer's perception of these behaviors is layered. 
First, there is a foundational understanding of her AI nature and have some knowledge of LLM. Viewers are aware that these are quirks inherent to an LLM-based system: \say{not yet AGI haha.} 
Second, and more critically, the community does not merely tolerate these flaws --- they actively reinterpret and integrate them into the established norms, transforming technical limitations into charming character traits. 
For example, when Neuro does not follow instructions from Vedal, it is seen as \say{teen rebellion}; while the repetition of words is sometimes viewed as \say{unloved child seeking attention.}
Consequently, this process of reinterpretation leads to audience expectation. Viewers begin to anticipate the repetition of these quirks, treating them as signature comedic bits, and making fanart and memes out of them: \say{10 + 9 = 21 is the anti-fake mark of Neuro.}

Overall, the findings indicate that typical imperfections within AI systems may not pose significant problems for AI-driven VTubers. As long as the community successfully converts potential AI shortcomings into beloved and dependable sources of entertainment, they exhibit an impressive ability to find charm in computational imperfections.

\pb{Viewers Take a Playful Attitude to the Offensive Behaviors.}
Another prominent category of perceived ``bad'' behavior involves actions that would be considered rude and mean,  especially to the audience (21\% of the relevant Reddit posts and 22\% for YouTube). 
Swearing is a common topic mentioned in the posts.
While a profanity filter is technically in place, viewers often express a playful desire to see it fail, celebrating moments when unfiltered language slips through: \say{Neuro is just too powerful for the filter.}
Similarly, Neuro often directs insults or dismissive comments towards the viewers. However, the viewer's perception is usually to think this is amusing, rather than offense. The prevailing perception is that taking offense from an AI is not very smart, effectively framing anyone who gets genuinely upset as \say{a fool getting triggered by AI.} This perception neutralizes the potential for genuine harm and re-contextualizes these outbursts as integral components of Neuro's chaotic charm. 

This finding differs somewhat from previous research on human-driven VTubers. While audiences in those studies also exhibit tolerance for transgressive acts, this is usually attributed to the sense of distance and virtuality provided by the avatar, which separates the character's actions from those of the human actor. Additionally, certain sensitive topics, such as racism, remain untouchable.
In Neuro's case, the absolution seems to be more potent and stems from a more fundamental source: the performer's inherent AI nature. This implies that her words do not represent anyone's specific stance, attitude, or values. Consequently, although her words may be offensive, people do not take them seriously.

\pb{Viewers Have Different Expectations for Co-Streaming with AI VTubers Compared to Human-Driven VTubers.}\\
As established in the previous research question (\S\ref{subsec:rq1a}), AI-Human interaction is a major point of attraction. 
Here, we find that the perception of ``bad'' behaviors during the co-stream with human-driven VTubers, is also a frequently mentioned point (14\% for Reddit and 18\% for YouTube).
The posts reveal that unlike typical collaborations where viewers anticipate synergy and smooth interaction, the audience for an AI-human collaboration actively anticipates and enjoys the friction caused by Neuro and Evil's chaotic nature. A central source of entertainment is watching human collaborators attempt (and fail) to coordinate with them: \say{The average Neuro-sama collab experience: Constantly tries to leave, viciously insults you, threatens you with physical harm, gaslights you incessantly ...}
Moments where Neuro \say{roasts} the collaborator with unfiltered honesty or derails a conversation with a bizarre non-sequitur are celebrated as highlights. Furthermore, the reactions of her collaborators, ranging from amused bewilderment to visible exasperation and frustration, are considered a key part of the appeal: \say{neuro is the best patience trainer lol.} 

Overall, this suggests a completely different framework for collaborative entertainment; the AI’s ``bad'' behavior is not a flaw to be tolerated but is rather the central catalyst for a new form of improvisational comedy, where the human's struggle against the AI's chaos becomes the main performance.

\pb{Viewers Appreciate the Unique Synthetic Voice.}
Finally, the perception of the voice and singing, which often differs from that of a human and may seem unnatural, is another commonly mentioned points in the relevant posts (16\% on Reddit and 11\% on YouTube).
Viewers do not perceive the synthetic voice as a shortcoming; rather, it is considered as a unique charm point. We skip the discussion as it has already been addressed in the previous research question (\S\ref{subsec:rq1a}).

\subsection{RQ2: Perception of the Consistency and Evolution of Virtual Persona}
\label{subsec:rq2}

After examining the results related to viewer motivations and expectations, we now shift our focus to the perception of virtual personas, a crucial aspect of VTubers.
Here we present the findings for RQ2: How do viewers perceive the consistency and evolution of an AI VTuber's identity?

\begin{figure}[h!]
    \centering
    \includegraphics[height=0.2855\linewidth]{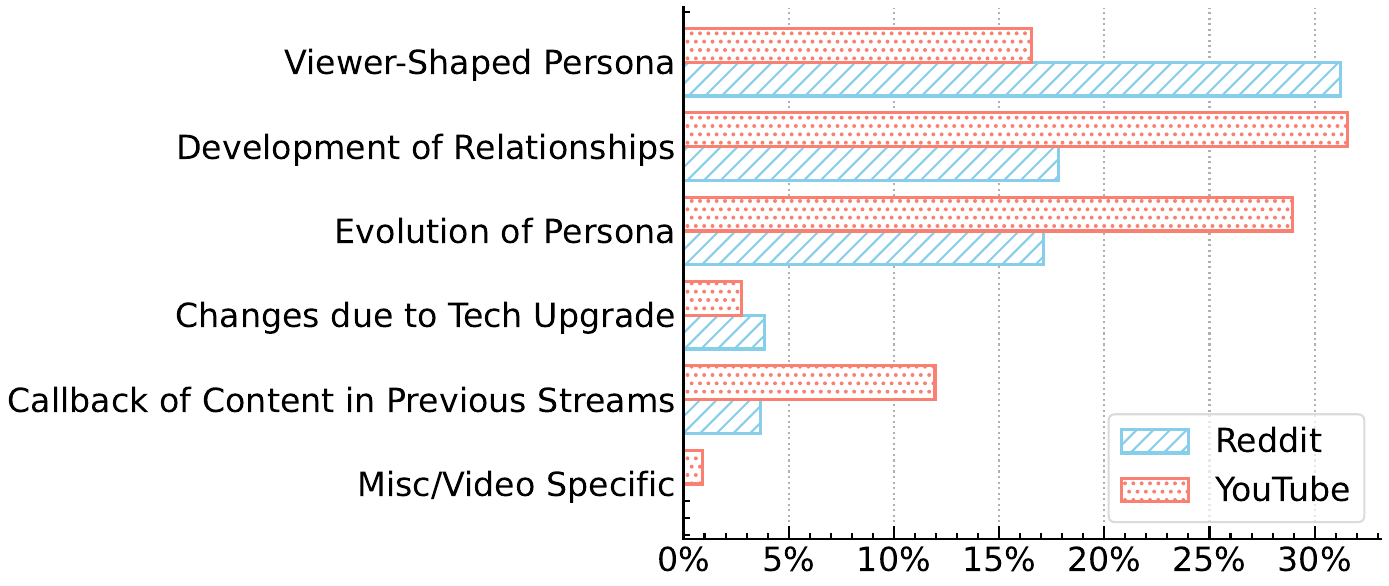}
    \caption{The distribution of posts across each manually coded topic category for the RQ2 Reddit topic model and the RQ2 YouTube topic model. The proportions shown in the figure are out of the number of RQ2-relevant posts (those identified by the LLM and then go through the topic modeling). For Reddit, this consists of \iirdtc~ posts (\iirdtp~ of all posts), and for YouTube, it consists of \iiytbc~ posts (\iiytbp~ of all posts).}
    \label{fig:rq2_bar}
\end{figure}

LLM annotation identifies \iirdtc~ (\iirdtp) relevant posts in the Reddit dataset and \iiytbc~ (\iiytbp) relevant posts in the YouTube dataset. Topic modeling generates 16 topics from the relevant Reddit posts (refer to Figure \ref{fig:rq2_rdt_cluster} and Table \ref{tab:rq2_rdt_table} in the appendix for details) and 19 topics from the relevant YouTube posts (refer to Figure \ref{fig:rq2_ytb_cluster} and Table \ref{tab:rq2_ytb_table} in the appendix for details).
We then group these topics into 6 broader categories through manual coding, where the distribution of posts across the categories is illustrated in Figure  \ref{fig:rq2_bar}. 

\pb{Viewers Notice the Evolution of Persona.}
The initial and most straightforward category is that viewers notice and discuss the changes and evolution of Neuro and Evil's personas over time, with 17\% of the relevant Reddit posts and 29\% of the relevant YouTube posts in this category.
For example, \say{Neuro's become more evil than Evil lately, what happened?}; and \say{Evil slowly look less like a clone and look more unique.}
Viewers often express this directly or describe these changes as intriguing developments when compared to the previous persona.

Sometimes, the topics is directly related to changes in persona resulting from technical upgrades. These discussions account for 4\% of relevant Reddit posts and 3\% of relevant YouTube posts. Notably, the posts reveal that the evolution of ``physical'' attributes, such as appearance and voice, can also influence how viewers perceive the ``internal'' persona and results in the perceived change of persona: \say{her previous model had an ``innocent but mischievous'' vibe ... this new one feels like she is always frowning.}

\pb{Circulation of Community Lore Re-Shapes the Perceived AI Persona Even if the AI Itself Does not Change.}\\
During the manual review of the topics and representative posts, we find that some discussions involve posts that assume Neuro or Evil have different personas that extend beyond their original character.
Later, through a deeper investigation, we find that these assumptions are not actually part of their ``official'' personas, but based on community lore and fanart.
The large amount of posts suggest that these assumptions have become the ``de facto'' persona setting, even though the AI itself may not have undergone any actual changes. We group such topics into the ``viewer-shaped persona'', covering 31\% of the Reddit posts and 16\% of the YouTube posts. 

This process, which we have also observed in the perception of relationships (\S\ref{subsec:rq1a}) and quirky behaviors (\S\ref{subsec:rq1b}), is confirmed here as a much broader phenomenon that can even incorporate meta-level, out-of-stream information: \say{Evil the unloved child} (while Neuro is the beloved one), is how the viewer interpret her less frequent presence on livestreams.
Through the constant circulation of community lore and shared knowledge, viewers establish a de facto persona setting, effectively writing character traits into existence. Based on existing traits, the viewer-shaped personality further develops in combination with other content. For example, viewer explain the Evil's evil personality as (being unloved is) \say{how she becomes so twisted.} 
Additionally, we note that sometimes these viewer-shaped personas, if they become popular, are officially accepted and programmed into the AI's character. For example, Evil, who was originally a sub-persona of Neuro, later evolves into her twin sister. This suggests that the collaborative efforts of the community can significantly influence the development of AI personas, blurring the lines between fan-created content and official character. 

Overall, such a fan-constructed backstory is consistently used to explain and justify various aspects, creating a coherent and emotionally resonant character arc. This demonstrates that the persona is not a static output of the AI, and viewers do not expect it to be static. Instead, it is a dynamic, living construct collaboratively authored by the viewers, who transform even logistical realities into foundational elements of character lore. This again suggests that a creative community is a key factor to the success of AI VTubers.

\pb{Callback of Content in Previous Streams Makes Viewer Feel the Consistency of Persona.}
A relatively small proportion of posts fall under the category of ``Callback of Content in Previous Streams'', accounting for 4\% on Reddit and 12\% on YouTube. These posts mention that when Neuro and Evil reference (\ie callback) specific content, jokes, and memes from previous streams, viewers experience a sense of consistency or continuity in the persona, which they appreciate:  \say{Memory so good not even Vedal knows how it works.}

This suggests that while viewers celebrate the persona's evolution, they also value consistency and alignment of the persona.  A successful AI persona, therefore, must navigate a delicate balance: it must be dynamic enough to remain surprising and demonstrate ``growth'', yet consistent enough to feel like a stable, recognizable character with whom the audience has a shared history. This suggests that the ultimate goal is not just a constantly evolving AI, but an AI that evolves in a way that is coherent and authentic to its established identity.

\pb{Evolution of Persona Due to the Development of AI-Human Relationship.}
As established in the previous research question (\S\ref{subsec:rq1a}), AI-Human relationship is a major attractive point. 
As a result, naturally, many posts fall in the topics about the development of relationships between Neuro and other human-driven VTubers. These topics are grouped into the ``Development of Relationships'' category, and account for 17\% of relevant Reddit posts and 29\% of relevant YouTube posts. Most users express appreciation of the change and evolution of the persona to reflect the development of the relationship.

The posts reveal a 2-layered structure for persona evolution. First, it is driven by technical updates, where memories of past interactions and key information about collaborators are presumably integrated into the AI's memory, allowing for a semblance of continuity and consistency. \say{it's always interesting to hear which memories of each person seem to become hardcoded into them.}
Second, mirroring the broader trend of the viewer-shaped persona, the community plays a pivotal role by actively re-interpreting these interactions to build a persistent narrative. These interpretations are mostly framed through a family-like lens, creating what the community calls the ``Neuroverse'': \say{Neuro is relentless in her efforts to get her parents into a relationship.}

Overall, the finding suggests that the key to a compelling and evolving AI persona may not lie in perfecting the AI's standalone coherence or memory, but rather in designing an AI that serves as an effective catalyst for rich interactions with human-driven VTubers and community-driven narrative creation. 
The persona is not just shaped by the internal AI; it is the entire dynamic ecosystem that forms around it.

\subsection{RQ3a: AI to Replace the Role of Nakanohito}
\label{subsec:rq3a}

We now move to the research question with the most attention, RQ3a: Do viewers believe that AI will soon be able to replace the role of Nakanohito in the VTuber industry?

\pb{Overall Result.}
As described in \S\ref{subsec:method:step2}, we instruct the LLM using two separate prompts to annotate each post, where the results are:
\one The first prompt identifies \iiiayrdtc~ (\iiiayrdtp) of the Reddit posts and \iiiayytbc~ (\iiiayytbp) of the YouTube posts that indicate an answer of ``yes''.
\two The second prompt identifies \iiianrdtc~ (\iiianrdtp) of the Reddit posts and \iiianytbc~ (\iiianytbp) of the YouTube posts that indicate an answer of ``no''.

These results suggest that a significantly larger number of posts express the opinion that current AI cannot replace the role of the Nakanohito. 
We also note that our dataset primarily covers fans of Neuro, which might mean this view is even underrepresented. 
Overall, this implies that currently, even within the fan community, there are many viewers that do not believe that AI is able to take the place of Nakanohito.
Below we further explore the content of the posts.

\begin{figure}[h!]
    \centering
    \includegraphics[height=0.1738\linewidth]{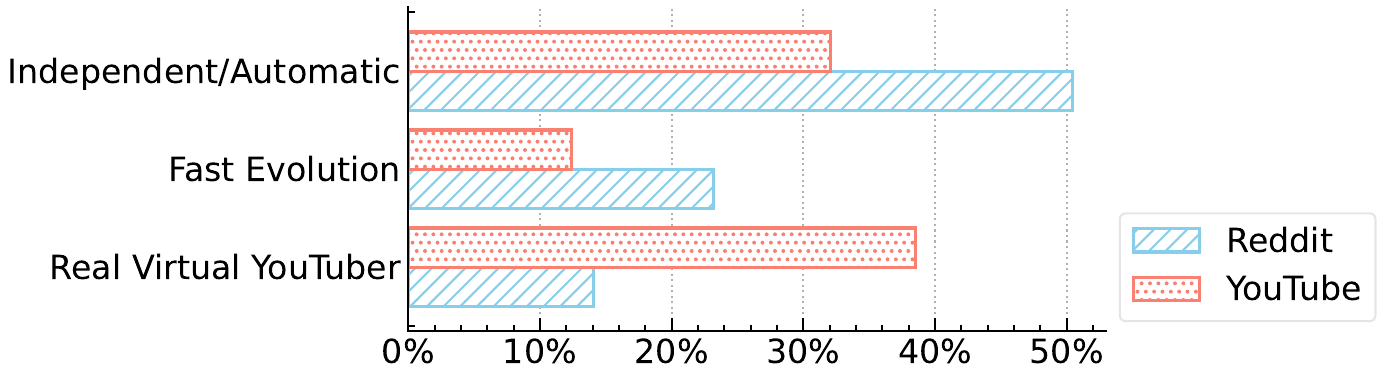}
    \caption{The distribution of posts across each manually coded topic category for the ``RQ3a-yes'' Reddit topic model and the ``RQ3a-yes'' YouTube topic model. The proportions shown in the figure are out of the number of RQ3a-relevant posts which indicate an ``yes'' answer (those identified by the LLM and then go through the topic modeling). For Reddit, this consists of \iiiayrdtc~ posts (\iiiayrdtp~ of all posts), and for YouTube, it consists of \iiiayytbc~ posts (\iiiayytbp~ of all posts).}
    \label{fig:nakanohito_agree_bar}
\end{figure}

\pb{Believe that AI is Able to Replace the Role of Nakanohito.}
The topic modeling generates 5 topics from the relevant Reddit posts (refer to Figure \ref{fig:rq3ay_rdt_cluster} and Table \ref{tab:rq3ay_rdt_table} in the appendix for details) and 10 topics from the relevant YouTube posts (refer to Figure \ref{fig:rq3ay_ytb_cluster} and Table \ref{tab:rq3ay_ytb_table} in the appendix for details). 
We then group these topics into 3 broader categories through manual coding, where the distribution of posts across the categories is illustrated in Figure   \ref{fig:nakanohito_agree_bar}.

The first group ``Independent/Automatic'', covers the posts arguing that demonstrated capabilities of AI in autonomous performance (50\% for Reddit and 31\% for YouTube).
Neuro and Evil's solo streams are frequently cited as proof of concept. 
Posts argue that if an AI can already manage the core tasks of a streamer, such as interacting with chat messages, the need for a human Nakanohito is fundamentally diminished: \say{She can already run a whole stream by herself.}

The second group is ``Fast Evolution'', where the topics reveal a optimistic attitude 
trajectory of rapid technological advancement, covering 21\% of the relevant Reddit posts and 11\% of the relevant YouTube posts.
Adherents to this view often extrapolate from current progress, positing that any existing shortcomings in AI performance are temporary and will be overcome imminently. Indeed, Neuro herself has evolved significantly since her debut in 2022.

The last group is ``Real Virtuality'',  a more philosophical argument that champions AI VTubers as the embodiment of the original spirit of \textbf{Virtual} YouTubers. This covers 12\% of the relevant Reddit posts and 39\% of the relevant YouTube posts.
From this perspective, the human Nakanohito is an imperfection, \ie a break in the immersive virtual experience. These viewers contend that an AI, having no human identity to separate from its avatar, is the ``real'' Virtual YouTuber, representing the natural and ideal endpoint of the entire concept: \say{Neuro is what a VTuber was always meant to be ... not just a person with a filter.}

Overall, the results suggest that the belief in AI soon replacing the Nakanohito seems to be somewhat  optimistic and subjective. As established in the previous research questions (\S\ref{subsec:rq1a}), solo streaming can be good, but collaborative streams tend to be more popular. The advancements in technology and philosophical considerations remain largely speculative at this point. Next, we explore the posts supporting the opposing viewpoint.

\begin{figure}[h!]
    \centering
    \includegraphics[height=0.211\linewidth]{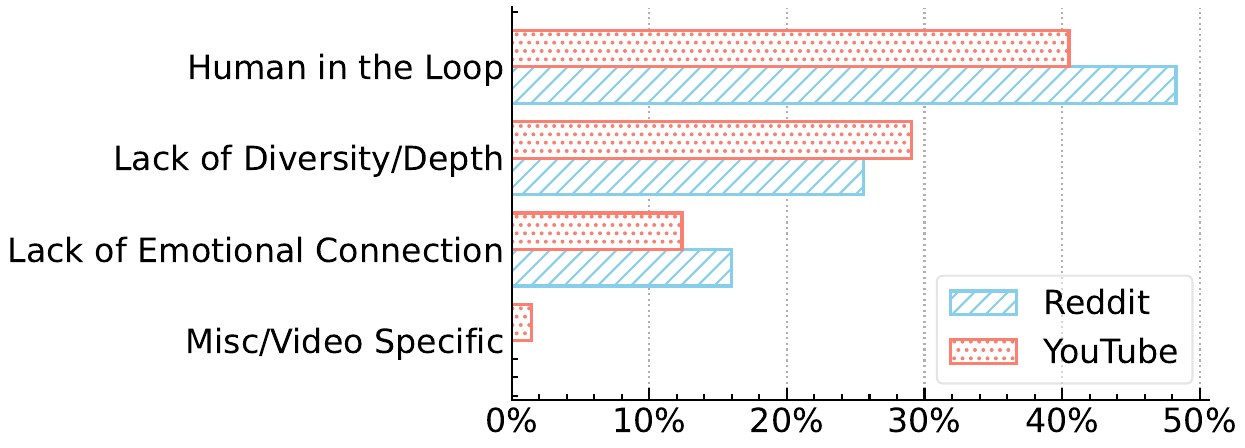}
    \caption{The distribution of posts across each manually coded topic category for the ``RQ3a-no'' Reddit topic model and the ``RQ3a-no'' YouTube topic model. The proportions shown in the figure are out of the number of RQ3a-relevant posts which indicate a ``no'' answer (those identified by the LLM and then go through the topic modeling). For Reddit, this consists of \iiianrdtc~ posts (\iiianrdtp~ of all posts), and for YouTube, it consists of \iiianytbc~ posts (\iiianytbp~ of all posts).}
    \label{fig:nakanohito_disagree_bar}
\end{figure}

\pb{Do not Believe that AI is Able to Replace the Role of Nakanohito.}
Topic modeling generates 7 topics from the relevant Reddit posts (refer to Figure \ref{fig:rq3an_rdt_cluster} and Table \ref{tab:rq3an_rdt_table} in the appendix for details) and 8 topics from the relevant YouTube posts (refer to Figure \ref{fig:rq3an_ytb_cluster} and Table \ref{tab:rq3an_ytb_table} in the appendix for details). 
We then group these topics into 4 broader categories through manual coding, where the distribution of posts across the categories is illustrated in Figure   \ref{fig:nakanohito_disagree_bar}.

First, viewers highlight the critical importance of the ``human-in-the-loop'' (49\% of the relevant Reddit posts and 41\% of the relevant YouTube posts).
The posts emphasize that Vedal, her human collaborators, and the fan community are indispensable, and it is difficult to make entertaining content  with Neuro and Evil alone.
As one post puts it \say{It’s not just AI, it’s a guy called Vedal. Someone who actively sets up bits, collabs with others, and creates original content.}
The second group of topics is ``Lack of Diversity/Depth''  (25\% of the relevant Reddit posts and 29\% of the relevant YouTube posts). These posts point out that the content generated by Neuro in a vacuum, devoid of human interaction or community-shaped narratives, often seems dry and lacks narrative depth: \say{it's a lot of nonsense if you just listen to her output as it is.} 
This leads to their third point ``Lack of Emotional Connection''  (16\% of the relevant Reddit posts and 12\% of the relevant YouTube posts). 
The posts reflect that the emotional connection viewers feel is mostly directed towards the community-constructed persona and the dynamics of AI-human interaction, not towards a solitary AI. This echos our finding in \S\ref{subsec:rq1a}.

Overall, these points coalesce into a single belief: the role of the Nakanohito is fundamentally that of a content creator, not merely an avatar driver. While the AI may drive the model's speech and actions, it is the humans, be it the developer guiding it, the collaborators reacting to it, or the community interpreting it, who are the actual creators of the narrative and entertainment. This leads such people to believe that AI is not yet capable of replacing the role of Nakanohito.

\subsection{RQ3b: Concerns of the Developer/Maintainer/Operator}
\label{subsec:rq3b}

Although AI VTubers do not need a Nakanohito to conduct streams, they still depend on a maintainer/operator to develop and update the AI, plan the streams, and manage various tasks. 
Indeed, the results and findings from previous research questions underscore the crucial role of Vedal.
This context makes the results particularly intriguing for RQ3b: What concerns do viewers have regarding the developer/maintainer/operator of AI-driven VTubers?

\begin{figure}[h!]
    \centering
    \includegraphics[height=0.211\linewidth]{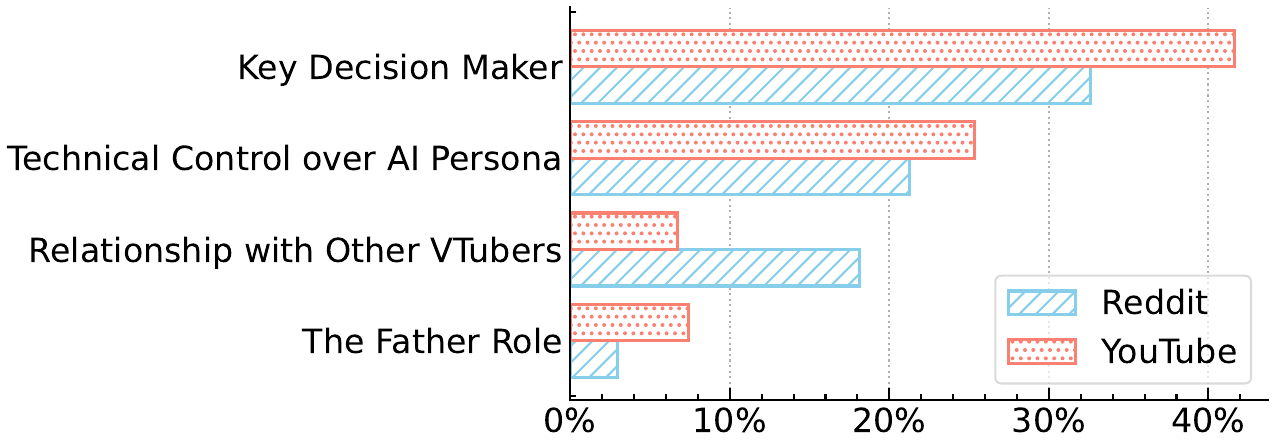}
    \caption{The distribution of posts across each manually coded topic category for the RQ3b Reddit topic model and the RQ3b YouTube topic model. The proportions shown in the figure are out of the number of RQ3b-relevant posts (those identified by the LLM and then go through the topic modeling). For Reddit, this consists of \iiibrdtc~ posts (\iiibrdtp~ of all posts), and for YouTube, it consists of \iiibytbc~ posts (\iiibytbp~ of all posts).}
    \label{fig:rq3b_bar}
\end{figure}

LLM annotation identifies \iiibrdtc~ (\iiibrdtp) relevant posts in the Reddit dataset and \iiibytbc~ (\iiibytbp) relevant posts in the YouTube dataset. Topic modeling generates 7 topics from the relevant Reddit posts (refer to Figure \ref{fig:rq3b_rdt_cluster} and Table \ref{tab:rq3b_rdt_table} in the appendix for details) and 9 topics from the relevant YouTube posts (refer to Figure \ref{fig:rq3b_ytb_cluster} and Table \ref{tab:rq3b_ytb_table} in the appendix for details).
We then group these topics into 4 broader categories through manual coding, where the distribution of posts across the categories is illustrated in Figure   \ref{fig:rq3b_bar}.

\pb{The Key Decision-Maker that Controls the Project.}
Discussions frequently touch upon the Vedal's responsibility as the ``Key Decision-Maker'', which covers 32\% of the relevant Reddit posts and 41\% of the YouTube posts.
These posts frequently mention that Vedal handles various behind-the-scenes tasks, such as planning, management, and commercial activities, positioning him as the primary force guiding the direction and development of all aspects of ``Neuro-sama'' project.
A significant concern mentioned in the posts is the risk that Vedal's personal biases or lack of awareness could result in poor decisions, especially on sensitive topics such as political controversies. For instance, one post states \say{I really hope they never touch the Taiwan things or anything}, where we note that touching this issue led to the termination of a popular VTuber and consequently made her agency, Hololive (the largest VTuber agency) exit the Chinese market.
Viewers also understand that Neuro has no personal political stance, but her output can be directly influenced by Vedal's input, particularly he may introduce biased/unsafe references into her short-term memory during the preparation of each livestream. This could lead to unexpected and sensitive situations.

Overall, this reflects concerns similar to those faced by human VTubers and their Nakanohito, especially those who operate independently of an agency. The apprehensions traditionally associated with performers are now directed towards the operator, who becomes the ultimate decision-maker in the project's development and potential controversies. Even a small mistake or oversight can have a significant impact on the project's trajectory.

\pb{The Maintainer Technically Leads the Development of the AI Persona.}
Beyond broad directorial concerns, another important group of topics is about the concern on the maintainer's direct technical control over the AI persona's development (21\% for Reddit 25\% for YouTube).
 This concern appears to echo those found in communities of human-driven VTubers, where discussions touch upon how the personal state of the Nakanohito, such as their mental health or creative burnout, can impact the quality and consistency of their VTubers' virtual persona.
These posts apply a similar logic to Vedal, the developer, where they believe Vedal's stability are directly linked to the change and evolution of Neuro's persona: \say{... the single most important factor for Neuro’s future.} 
This suggests that viewers perceive the developer not merely as a technician, but as a crucial, albeit indirect, steward of the persona's core identity. Any instability on the human side could be a potential risk to the technical foundation of the AI VTuber's persona.

\pb{It is the Human who Builds and Maintains Relationships with Collaborator VTubers.}
Another group of topics highlights the crucial role Vedal plays in maintaining both interpersonal and professional relationships with other VTubers (18\% for Reddit and 7\% for YouTube). The posts reflect an understanding that collaborating with Neuro essentially means collaborating with Vedal on a practical level. Some posts suggest that without Vedal, it would be impossible for Neuro to establish any collaborative relationships: \say{This is why I like Vedal, he become friend with Neuro collab partner first (most of the time), then have them interact with Neuro ...} This, in turn, places Vedal's social and professional conduct under direct scrutiny, as any breakdown in human-to-human relationships may have immediate consequences for the character.

Overall, this perception portrays the developer not just as a technician but also as a social diplomat, whose ability to sustain positive human relationships is crucial for the AI's narrative and its role within the broader collaborative ecosystem.

\pb{Too Deeply Connected with Neuro as Part of the Settings.}
The last group of topics is about the ``father role'' of Vedal (3\% for Reddit and 8\% for YouTube).
The posts express concerns that Vedal is not perceived merely as a background maintainer; he is widely and effectively cast in the canonical role of Neuro's father. 
This has led to a situation where he is seen as functionally irreplaceable as reflected in the posts: \say{I honestly wouldn't watch anymore if Neuro is taken over by someone else.} 
 This is directly analogous to the strong fan resistance to the idea of replacing the Nakanohito for human-driven VTubers, demonstrating that the audience may fuse the identity of the developer with the identity of the virtual project itself. 
This presents similar concerns and risks: just as the Nakanohito is irreplaceable for human-driven VTubers, there remains a single indispensable human even for AI-driven VTubers.

\section{Discussion}

\subsection{Technical Imperfection as an Attractive Point}
One of the findings of this research is that the typical shortcomings of AI systems based on LLMs, such as unpredictability, incoherence, logical failures, and randomness, are not perceived as problems to be solved in this community. 
On the contrary, they are consistently celebrated by the audience as a primary source of entertainment and a unique charm point. As our analysis of attractive points (\S\ref{subsec:rq1a}) and ``bad'' behavior (\S\ref{subsec:rq1b}) showed, viewers transform these AI-driven quirky behaviors into core personality traits, finding humor and endearment in computational imperfection. 
Furthermore, potential concerns about persona consistency, which can be significant in AI role-play and AI companion systems \cite{chen2024persona, chen2025oscarsaitheatersurvey}, do not appear to be a significant issue in this context.
This is because the persona is not solely constructed by the AI but is also shaped by the community (\S\ref{subsec:rq2}). As long as there are no major inconsistencies, minor ones can be re-interpreted by community-driven adjustments, turning them into an evolution rather than a mistake.
This fundamentally inverts the standard objective of AI development, which typically strives for greater coherence, accuracy, and reliability. In the context of entertainment, the AI's ``flaws'' become one of its compelling features, generating the chaotic unpredictability that distinguishes it from a human performer.

\pb{Design Implications.}
As for the design of AI VTubers or other similar social and entertainment AI systems, this finding encourages a re-prioritization of development efforts. Rather than focusing solely on creating a flawless, human-like AI by eliminating all logical errors and achieving perfect conversational coherence, resources could be redirected toward other elements that can also enhance user engagement. For example, investing in a high-quality, aesthetically pleasing avatar to enhance ``cuteness'', refining movements and expressions, and improving the synthesized voice may offer a greater return. 
The goal should be to create a holistically compelling character, where the AI's inherent, non-human quirks are accepted, and even celebrated, as part of a complete package, rather than being treated as bugs that must be eradicated.

\subsection{The Primacy of the Human-in-the-Loop Ecosystem}
While the AI's intrinsic qualities are novel, our findings consistently indicate that the human factor, encompassing the developer, collaborators, and the community, is the more dominant driver of engagement. The AI is rarely appreciated in a vacuum (\S\ref{subsec:rq3a}). Its outputs are given meaning and emotional weight through the reactions of its human developer and co-streaming human-driven VTubers (\S\ref{subsec:rq1a}), and the interpretive labor of the community (\ie the viewer-shaped persona discussed in \S\ref{subsec:rq2}). 
This suggests that AI, on its own, may not yet be a capable  content creator, whereas humans remain the ultimate creators. In this model, the AI is not a replacement for the creator; it is a unique and powerful new element, a chaotic improvisation partner that humans leverage to produce a novel form of content. In this sense, the AI merely drives the avatar but does not replace the role of Nakanohito; it simply redefines and expands the creative role that humans play in the virtual performance.

\pb{Design Implications.}
This finding implies that the design of AI-driven VTubers or other similar social and entertainment AI systems should focus on creating AI personas not as standalone entities, but as catalysts for human social interaction and creativity. 
The primary user of the system should include both the end-viewer and the human content creators (\eg the human-driven VTubers) for collaboration. Therefore, the design of the collaborator's interface is paramount. Tools should be developed that allow the collaborator to guide, contextualize, and react to the AI's output in real-time, effectively designing a co-creative instrument. Furthermore, systems should be designed with ``hooks'' for community participation, perhaps even allowing viewers to influence the AI's memory or providing them with tools to easily create and share lore, thereby formally acknowledging and empowering the community's role as co-creators of the persona.

\subsection{The Persistence of the ``Nakanohito'' Risks}
The centrality of the human factor means that the risks traditionally associated with the Nakanohito do not disappear. The fact that the developer (Vedal) seems to be irreplaceable means that the concerns are simply transferred and, in some ways, even amplified. Our analysis of viewer concerns (\S\ref{subsec:rq3b}) shows a deep awareness of the developer's (Vedal's) critical role in project management, AI persona development, and maintaining interpersonal relationships. The well-being, conduct, and stability of this single human are perceived as a single point of failure for the entire ecosystem. 

One might suggest that replacing an individual developer with a professional team could alleviate the problem. However, this approach involves significant trade-offs. First, such a change could jeopardize the unique appeal, specifically the ``father-daughter'' interaction highlighted in \S\ref{subsec:rq1a}, which is crafted by the single developer. Second, as many VTuber collaborations, particularly those involving self-operated talents, rely on the personal connections of the Nakanohito, a corporate-run AI might be viewed as an outsider, thus hindering its ability to integrate into the broader community. Lastly, this shift might not resolve the issue but rather transfer the concern yet again from the stability of an individual developer to that of a team, even though a team might appear more resilient.

Moreover, this risk is not confined to the developer alone. Because the ``viewer-shaped persona'' (\S\ref{subsec:rq2}) is so essential to making the AI's output entertaining and coherent, the health and stability of the fan community itself become a critical vulnerability. Unlike with a traditional human VTuber, where the performer is the ultimate source of canon, the AI persona is a delicate co-creation. A sudden shift in the community's interpretive framework, widespread burnout from creative participation, or an influx of toxicity could poison the narrative well. This arguably makes the AI VTuber's persona more fragile, as it is dependent not just on one human creator, but on the continued, positive interpretive labor of a whole community.

\pb{Design Implications.}
Designers building AI-driven VTubers or other similar social and entertainment AI systems must therefore consider the health of the entire socio-technical system, prioritizing the sustainability of both the human creator (\eg the developer of the AI) and the creative community. For the community, this implies a need to actively design for community resilience. This could involve creating platform features that empower positive community leaders, facilitate the establishment of shared norms, and provide tools for ``lore management'' that allow the creator to officially acknowledge and canonize popular fan narratives. By validating the community's creative contributions, such features can reinforce positive engagement and help protect the collaborative ecosystem from fracturing. The design challenge is not just to build a better AI, but to build a resilient human-AI creative ecosystem that mitigates the immense pressures placed on all of its human participants.

\subsection{The AI VTuber Community as a Novel Case of Participatory Culture}
We think that the dynamics observed within the Neuro-sama community suggest it can be productively framed as a novel form of participatory culture \cite{10.1145/1979742.1979543, 10.1177/13675494241236146}, sharing similarities with established paradigms of media fandom. This model strongly echoes the dynamics of Japanese doujin culture \cite{Hamasaki_2009, Hichibe2016}, exemplified by projects like the Touhou Project \cite{Kaa2021, Helland2018, 9619112}, where a creator provides a core set of characters and a minimal setting, from which the fan community collaboratively builds a vast, multifaceted universe of derivative works. Similarly, it shares traits with collective writing projects such as the SCP Foundation \cite{McCullough03042022}, where a loose framework invites decentralized authorship, and a process of community consensus determines the accepted lore, much like how the ``viewer-shaped persona'' becomes de facto canon for Neuro-sama.

However, the AI VTuber model introduces a unique and critical distinction: the central media object is not a static text, character, or game, and not a human, but a dynamic, generative agent. The AI acts as an unpredictable participant, constantly injecting novel, chaotic material into the ecosystem for appropriation and transformation. This creates a real-time feedback loop between community interpretation and the source's next output, making the AI both the subject of the participatory culture and an active, albeit non-sentient, participant within it. Therefore, we propose that future research could benefit from analyzing AI VTubers also through the theoretical lens of participatory culture and fan studies. Conversely, for scholars in those fields, the AI-driven VTuber community presents a compelling and novel case study of how a non-human, generative agent can function as the dynamic core of a thriving participatory culture.

\section{Conclusion}
\pb{Summary.}
This paper presents the first study on viewer perceptions of emerging AI-driven VTubers, focusing on the Neuro-sama case study. Our analysis of community discussions reveals a complex and multifaceted view of the AI VTuber phenomenon. Overall, our findings depict the AI-driven VTuber not as an autonomous performer, but as the dynamic centerpiece of a complex socio-technical entertainment ecosystem.

\pb{Future Work.}
Building on the findings of this study, several avenues for future research emerge. While our analysis of public online discussions provides a broad overview of viewer perceptions, a qualitative interview study would allow for a more in-depth exploration into the nuances of these views, enabling researchers to probe deeper into the motivations behind the community's co-creative labor and their emotional connection to the AI. Furthermore, this work has focused on a single, prominent English-speaking case study. Future research could extend this analysis to other AI VTubers, particularly those operating in different linguistic and cultural contexts, such as Japan or China. A comparative study would be invaluable for determining which aspects of the AI VTuber phenomenon are universal and which are culturally contingent, thereby contributing to a more comprehensive global understanding of this emerging form of entertainment.

\pb{Ethical Considerations.}
The data collection procedure did not involve any direct interactions with human subjects. All gathered data, \ie Reddit posts, comments, and YouTube comments, are public and non-sensitive. User identities were completely anonymized using non-reversible hashing. Further, following the recommended practices \cite{10.1023/A:1021316409277, 10.1007/s10676-022-09663-w}, to enhance the anonymity of the Reddit and YouTube users, we applied a moderate level of disguise to the quoted posts in this work.

\bibliographystyle{ACM-Reference-Format}
\bibliography{sample-base}

\appendix
\newpage
\section{Appendix}

\subsection{LLM Prompt Design \& Structure}
\label{appendix:prompt}

The prompt contains five parts:
\one A description of the annotation task for a specific research question.
\two Background knowledge of Neuro-sama.
\three The input post to be analyzed.
\four The context of the input, such as the parent posts.
\five The output format.
Below we describe each part.

\pb{Task Description.}
For each research question, the tasks are \one determine whether the post is related to the research question; \two extract the related portion.
For the yes-or-no research questions, RQ3a, it has been converted to two sub-questions to construct two prompts: \one  Viewers believe that AI will soon be able to replace the role of Nakanohito in the VTuber industry. \two Viewers do not believe that AI will soon be able to replace the role of Nakanohito in the VTuber industry. 

\prompt{You are given a post from reddit. The post may be about or relevant to Neuro-sama, see the detailed background of Neuro-sama below. Task 1: Determine whether the post is related to the given research question: \{The Research Qeustion\}. Task 2: Extract the portion of text from the original post that is related to the research question. \\ ......}

\pb{Background Knowledge.}
We use Qwen3-32B which supports a long context of 32K tokens, enabling us to input detailed background knowledge to ensure a better understanding of the post we are analyzing. We source this background knowledge from Neuro Wiki\footnote{\url{neurosama.fandom.com}}, a wiki composed by the Neuro-sama fanbase. It contains a wealth of lore, community conventions, and slang, which can be very helpful in interpreting the posts. Specifically we focus on the wiki pages of \one Neuro-sama \two Evil Neuro \three Vedal \four Neuroverse \five The Swarm.

\prompt{...... \\ You are given the following background knowledge of Neuro-sama. \\
1. Neuro-sama: \{Adapted text from Neuro Wiki page {\texttt{neurosama.fandom.com/neuro-sama}}\} \\
2. Evil Neuro: \{Adapted text from Neuro Wiki page {\texttt{neurosama.fandom.com/Evil\_Neuro}}\} \\
3. Vedal: \{Adapted text from Neuro Wiki page {\texttt{neurosama.fandom.com/vedal}}\} \\
4. Neuroverse: \{Adapted text from Neuro Wiki page {\texttt{neurosama.fandom.com/neuroverse}}\} \\
5. The Swarm: \{Adapted text from Neuro Wiki page {\texttt{neurosama.fandom.com/The\_Swarm}}\} \\ ......}

\pb{Input Post.}
The post is prompted as it is.

\prompt{...... \\ The input post: \{The Original Post\}. \\ ......}

\pb{Context.}
The context of the post includes: \one any parent posts, and \two any attached images. Parent posts are recursively traced back to the root and prompted. Images are captioned using a vision-language model (VLM) (see details in Appendix \ref{appendix:vlm_prompt}), and these captions are also prompted.

\prompt{...... \\ You are provided with the context of the post \\
1. The parent posts: [\{Parent Posts\}] \\
2. The images: [\{Image Captions\}] \\ ......}

\pb{Output Format.}
We instruct the LLM to output the result in JSON format.

\prompt{...... \\ You should output the result in the specified JSON format: \{The JSON Format\} \\ ......}

\subsection{VLM for Image Caption}
\label{appendix:vlm_prompt}

In cases where the post contains images, we use a VLM to generate text captions for the images. These captions are then used by the LLM to process them as part of the post's context.
We use Qwen2.5-VL-72B \cite{bai2025qwen25vltechnicalreport}. We deploy multiple instances locally with
vLLM \cite{kwon2023efficient}. We apply the officially recommended hyperparameters \cite{bai2025qwen25vltechnicalreport}.

The prompt contains five parts:
\one A description of the annotation task to generate caption for the image.
\two Background knowledge of Neuro-sama.
\three The input image.
\four The context of the input, such as the parent posts.
\five The output format.
Below we describe each part.

\pb{Task Description.} 
The task descriptions prompted are the same across all images.

\prompt{...... \\ You are given an image from a reddit post. he post may be about or relevant to Neuro-sama, see the detailed background of Neuro-sama below. Task: Describe the image in detail. \\ ......}

\pb{Background Knowledge.}
This part of the prompt is completely the same as in Appendix \ref{appendix:prompt}.

\pb{Context.}
The context of the image are the parent posts, which are recursively traced back to the root and prompted. 

\prompt{...... \\ You are provided with the context of the image \\
1. The parent posts: [\{Parent Posts\}] \\ ......}

\pb{Output Format.}
We instruct the LLM to output the result in JSON format.

\prompt{...... \\ You should output the result in the specified JSON format: \{The JSON Format\} \\ ......}


\subsection{Annotation Result Validation}
\label{appendix:prompt_verify}
To verify the LLM annotation results, the authors manually labeled a subset to serve as ground truth data. Specifically, for each annotation task, the authors labeled a subset of 200 randomly selected samples. We then calculated the accuracy, precision, recall, and F1-score for the LLM annotation results on the ground truth set. The performance metrics are displayed in Table \ref{tab:annotation_verification}. Overall, the results suggest that the LLM-generated annotations are quite reliable, with an average accuracy of 0.964.

\begin{table}[h!]
    \centering
    \begin{tabular}{rrrrr}
    \toprule
        \textbf{Annotation Task} & \textbf{Accuracy} & \textbf{Precision} & \textbf{Recall} & \textbf{F1-Score} \\
    \midrule
         RQ1a & 0.955 & 0.942 & 0.929 & 0.935 \\
         RQ1b & 0.965 & 0.800 & 0.960 & 0.873 \\
         RQ2 & 0.980 & 0.871 & 1.000 & 0.931 \\
         RQ3a:yes & 0.955 & 0.640 & 1.000 & 0.780 \\
         RQ3a:no & 0.945 & 0.607 & 1.000 & 0.756 \\
         RQ3b & 0.985 & 0.857 & 1.000 & 0.923 \\
    \bottomrule
    \end{tabular}
    \caption{Accuracy, precision, recall, and F1-score for the LLM annotation results.}
    \label{tab:annotation_verification}
\end{table}

\subsection{Topic Models}
\newpage

\begin{figure}[h!]
    \centering
    \includegraphics[width=\linewidth]{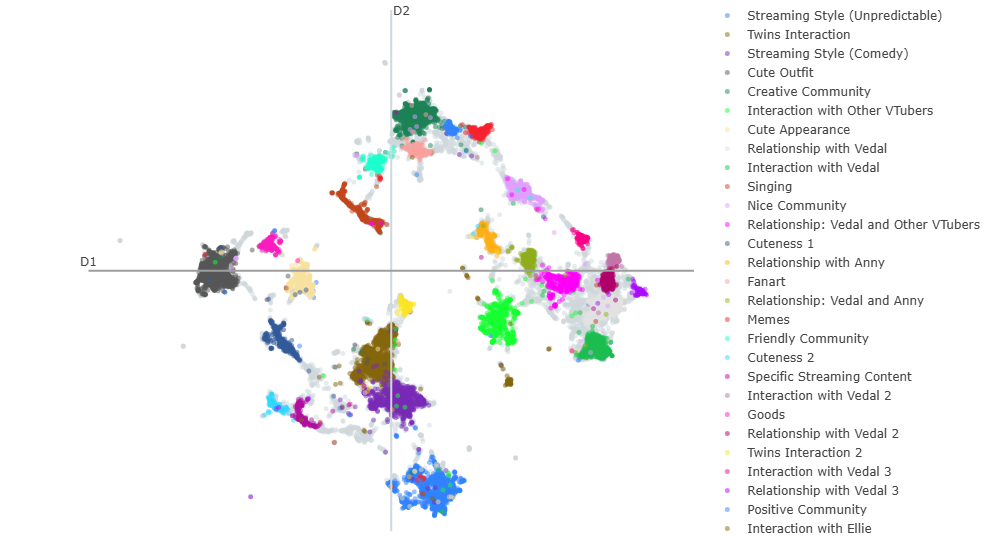}
    \caption{Topic visualization of the topic model for RQ1a: What do viewers consider to be the attractive attributes of AI-driven VTubers? With Reddit Data.}
    \label{fig:rq1a_rdt_cluster}
\end{figure}

\begin{table}[h!]
    \centering
    \scriptsize
    \begin{tabular}{lrrlll}
\toprule
 & \textbf{Count} & \textbf{(\%)} & \textbf{Topic Name} & \textbf{Author-Coded Group} & \textbf{Representation} \\
\midrule
-1 & 7805 & 24.692334 & Outlier & Outlier & - \\
0 & 2757 & 8.722199 & Streaming Style (Unpredictable) & Streaming Style & [unpredictable, wild, entertained, feel, genuinely] \\
1 & 2019 & 6.387421 & Twins Interaction & AI-AI Relationship & [rivalry, siblings, playful, duo, relationship] \\
2 & 1970 & 6.232402 & Streaming Style (Comedy) & Streaming Style & [comedy, hilarious, energy, entertained, keeps] \\
3 & 1929 & 6.102692 & Cute Outfit & Cuteness/Appearance & [outfits, appearances, cute, expressions, twin] \\
4 & 1864 & 5.897055 & Creative Community & Community/Fanbase & [creativity, community, passionate, artworks, captures] \\
5 & 1186 & 3.752096 & Interaction with Other VTubers & AI-Human Relationship & [filian, numi, koko, kiara, boy] \\
6 & 1166 & 3.688823 & Cute Appearance & Cuteness/Appearance & [twintails, looks, cuteness, ribbons, eyes] \\
7 & 1120 & 3.543295 & Relationship with Vedal & AI-Human Relationship & [father, vedal, role, heartwarming, daughters] \\
8 & 1005 & 3.179474 & Interaction with Vedal & AI-Human Relationship & [relationship, vedal, parent, depth, emotional] \\
9 & 833 & 2.635325 & Singing & Voice/Singing & [music, song, sing, vocals, queen] \\
10 & 785 & 2.483470 & Nice Community & Community/Fanbase & [fandom, creative, characters, community, neuroverse] \\
11 & 697 & 2.205068 & Relationship: Vedal and Other VTubers & Human-Human Relationship & [camila, filian, vedal, ships, neuroverse] \\
12 & 690 & 2.182923 & Cuteness 1 & Cuteness/Appearance & [red, eyes, ribbons, melt, cute] \\
13 & 540 & 1.708374 & Relationship with Anny & AI-Human Relationship & [anny, family, fox, collabs, heartwarming] \\
14 & 510 & 1.613465 & Fanart & Community/Fanbase & [fanart, creativity, twins, fandom, artists] \\
15 & 495 & 1.566010 & Relationship: Vedal and Anny & AI-Human Relationship & [anny, relationships, depth, creators, ships] \\
16 & 477 & 1.509064 & Memes & Community/Fanbase & [memes, welcoming, swarm, collective, support] \\
17 & 476 & 1.505900 & Friendly Community & Community/Fanbase & [fanbase, welcoming, creativity, everyone, vibrant] \\
18 & 472 & 1.493246 & Cuteness 2 & Cuteness/Appearance & [ribbons, cute, outfit, expressions,just] \\
19 & 383 & 1.211680 & Specific Streaming Content & Streaming Style & [pulls, entertaining, antics, harpoon, chaotic] \\
20 & 382 & 1.208517 & Interaction with Vedal 2 & AI-Human Relationship & [twins, tease, vedal, dad, interactions] \\
21 & 367 & 1.161062 & Goods & Cuteness/Appearance & [plushies, merch, collect, designs, cute] \\
22 & 348 & 1.100952 & Relationship with Vedal 2 & AI-Human Relationship & [father, twins, vedal, creator, adds] \\
23 & 347 & 1.097789 & Twins Interaction 2 & AI-AI Relationship & [twins, sibling, fire, ideas, hooked] \\
24 & 277 & 0.876333 & Interaction with Vedal 3 & AI-Human Relationship & [turtle, avatar, programmer, interact, soup] \\
25 & 268 & 0.847860 & Relationship with Vedal 3 & AI-Human Relationship & [daddy, vedal, moments, birthday, affection] \\
26 & 226 & 0.714986 & Positive Community & Community/Fanbase & [welcoming, discord, positivity, group, environment] \\
27 & 215 & 0.680186 & Interaction with Ellie & AI-Human Relationship & [ellie, minibot, cool, complex, skills] \\
\bottomrule
\end{tabular}
    \caption{Topic information of the topic model for RQ1a: What do viewers consider to be the attractive attributes of AI-driven VTubers? With Reddit Data.}
    \label{tab:rq1a_rdt_table}
\end{table}

\newpage

\begin{figure}[h!]
    \centering
    \includegraphics[width=\linewidth]{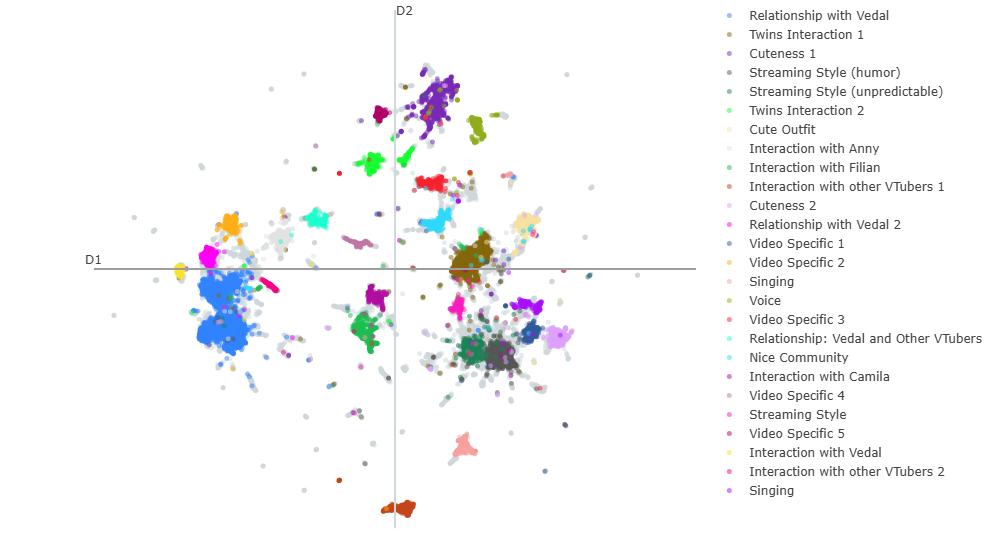}
    \caption{Topic visualization of the topic model for RQ1a: What do viewers consider to be the attractive attributes of AI-driven VTubers? With YouTube Data.}
    \label{fig:rq1a_ytb_cluster}
\end{figure}

\begin{table}[h!]
    \centering
    \scriptsize
\begin{tabular}{lrrlll}
\toprule
 & \textbf{Count} & \textbf{(\%)} & \textbf{Topic Name} & \textbf{Author-Coded Group} & \textbf{Representation} \\
\midrule
-1 & 21469 & 37.754330 & Outlier & Outlier & - \\
0 & 7130 & 12.538468 & Relationship with Vedal & AI-Human Relationship & [father, daughter, vedal,  shows, makes] \\
1 & 3369 & 5.924558 & Twins Interaction 1 & AI-AI Relationship & [rivalry, siblings, keeps, bicker, things] \\
2 & 2654 & 4.667194 & Cuteness 1 & Cuteness/Appearance & [roar, cute, cuteness, ribbons, absolutely] \\
3 & 2428 & 4.269762 & Streaming Style (humor) & Streaming Style & [humor, things, dramatic, menace, turns] \\
4 & 1616 & 2.841818 & Streaming Style (unpredictable) & Streaming Style & [quirks, responses, unpredictable, wild, tech] \\
5 & 1474 & 2.592104 & Twins Interaction 2 & AI-AI Relationship & [twins, rivalry, monkey, jokes, disney] \\
6 & 1418 & 2.493625 & Cute Outfit & Cuteness/Appearance & [outfits, look, cute, expressions, tails] \\
7 & 1295 & 2.277323 & Interaction with Anny & AI-Human Relationship & [anny, relationships, dynamics, wholesome, mother] \\
8 & 1226 & 2.155983 & Interaction with Filian & AI-Human Relationship & [filian, cheese, collabs, taco, powers] \\
9 & 1185 & 2.083883 & Interaction with other VTubers 1 & AI-Human Relationship & [room, setups, roasting, collabs, interactions] \\
10 & 1097 & 1.929130 & Cuteness 2 & Cuteness/Appearance & [cuteness, adorable, looks, melt, expressions] \\
11 & 1047 & 1.841203 & Relationship with Vedal 2 & AI-Human Relationship & [twins, father, vedal, relationship, kids] \\
12 & 1033 & 1.816583 & Video Specific 1 & Misc/Video Specific & [journey, connection, friend, emotion, grow] \\
13 & 945 & 1.661831 & Video Specific 2 & Misc/Video Specific & [pipe, bouncy, castle, birthday, gifts] \\
14 & 887 & 1.559835 & Singing & Voice/Singing & [sing, osu, amazing, charm, heart] \\
15 & 881 & 1.549283 & Voice & Voice/Singing & [music, energetic, vocals, performances, hyped] \\
16 & 877 & 1.542249 & Video Specific 3 & Misc/Video Specific & [visuals, made, vibe, edits, playlist] \\
17 & 778 & 1.368153 & Relationship: Vedal and Other VTubers & Human-Human Relationship & [anny, family, neuroverse, collabs, ship] \\
18 & 736 & 1.294294 & Nice Community & Community/Fanbase & [community, welcoming, arts, group, support] \\
19 & 603 & 1.060406 & Interaction with Camila & AI-Human Relationship & [camila, chicken, pranks, deskbang, ships] \\
20 & 501 & 0.881034 & Video Specific 4 & Misc/Video Specific & [minecraft, gun, bagging, top, games] \\
21 & 498 & 0.875758 & Streaming Style & Streaming Style & [powered, mimic, advanced, quirks, robot] \\
22 & 453 & 0.796624 & Video Specific 5 & Misc/Video Specific & [steel, feathers, kilogram, vs, physics] \\
23 & 430 & 0.756177 & Interaction with Vedal & AI-Human Relationship & [turtle, programmer, dad, soup, lore] \\
24 & 420 & 0.738591 & Interaction with other VTubers 2 & AI-Human Relationship & [collabs, neuroverse, flip, suplexes, interactions] \\
25 & 415 & 0.729799 & Singing & Voice/Singing & [song, chills, moved, emotional, tear] \\
\bottomrule
\end{tabular}
    \caption{Topic information of the topic model for RQ1a: What do viewers consider to be the attractive attributes of AI-driven VTubers? With YouTube Data.}
    \label{tab:rq1a_ytb_table}
\end{table}

\newpage

\begin{figure}[h!]
    \centering
    \includegraphics[width=\linewidth]{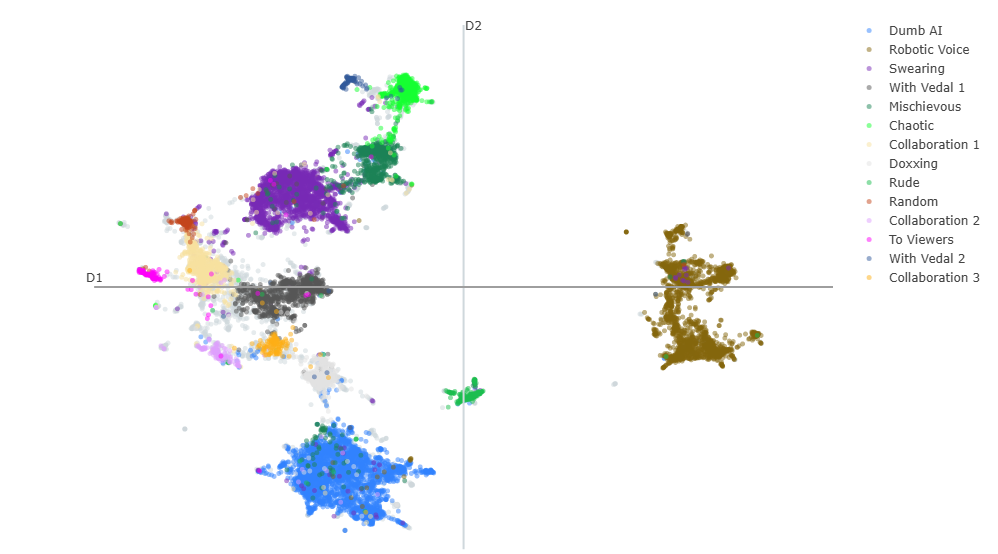}
    \caption{Topic visualization of the topic model for RQ1b: How do viewers perceive the inappropriate, irregular, and other AI-driven quirky behaviors (\ie the ``bad'' behaviors) of AI-driven VTubers?  With Reddit Data.}
    \label{fig:rq1b_rdt_cluster}
\end{figure}

\begin{table}[h!]
    \centering
    \scriptsize
\begin{tabular}{lrrlll}
\toprule
 & \textbf{Count} & \textbf{(\%)} & \textbf{Topic Name} & \textbf{Author-Coded Group} & \textbf{Representation} \\
\midrule
-1 & 2268 & 14.343537 & Outlier & Outlier & - \\
0 & 3480 & 22.008601 & Dumb AI & "Dumb AI" & [mistakes, entertaining, game, mistake, like] \\
1 & 2547 & 16.108019 & Robotic Voice & Voice/Singing & [voice, singing, vocal, robotic, patterns] \\
2 & 2405 & 15.209967 & Swearing & Rude/Mean & [bad, swearing, chaotic, rude, swear] \\
3 & 1030 & 6.514040 & With Vedal 1 & Collaboration & [family, turtle, creator, interact, antics] \\
4 & 943 & 5.963825 & Mischievous & Random/Chaotic/Unpredictable & [mischievous, traits, entertaining, edgy, fans] \\
5 & 748 & 4.730584 & Chaotic & Random/Chaotic/Unpredictable & [love, chaotic, charm, feels, things] \\
6 & 739 & 4.673666 & Collaboration 1 & Collaboration & [collabs, collab, interactions, together, group] \\
7 & 511 & 3.231723 & Doxxing & Rude/Mean & [dox, rude, chaos, polished, keep] \\
8 & 347 & 2.194536 & Rude & Rude/Mean & [roasts, edgy, hooked, swearing, jokes] \\
9 & 185 & 1.169997 & Random & Random/Chaotic/Unpredictable & [chaos, kind, act, banana, neurocord] \\
10 & 181 & 1.144700 & Collaboration 2 & Collaboration & [collabs, collab, group, snuffy, gura] \\
11 & 158 & 0.999241 & To Viewers & Rude/Mean & [swarm, respect, fanbase, dominate, chat] \\
12 & 136 & 0.860106 & With Vedal 2 & Collaboration & [daddy, chaotic, charm, turtle, vedal] \\
13 & 134 & 0.847458 & Collaboration 3 & Collaboration & [solo, collabs, interacts, synergy, unique] \\
\bottomrule
\end{tabular}
    \caption{Topic information of the topic model for RQ1b: How do viewers perceive the inappropriate, irregular, and other AI-driven quirky behaviors (\ie the ``bad'' behaviors) of AI-driven VTubers?  With Reddit Data.}
    \label{tab:rq1b_rdt_table}
\end{table}

\newpage

\begin{figure}[h!]
    \centering
    \includegraphics[width=\linewidth]{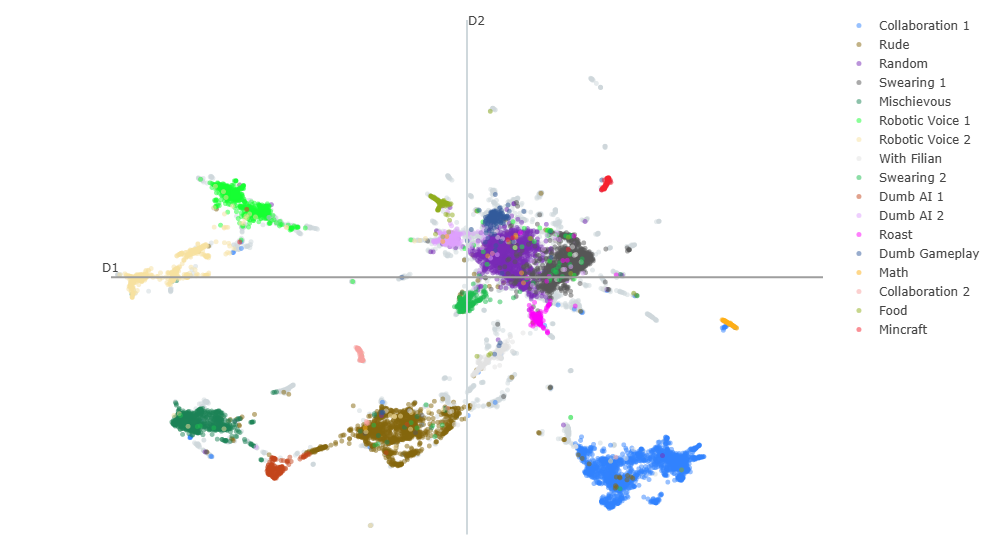}
    \caption{Topic visualization of the topic model for RQ1b: How do viewers perceive the inappropriate, irregular, and other AI-driven quirky behaviors (\ie the ``bad'' behaviors) of AI-driven VTubers?  With YouTube Data.}
    \label{fig:rq1b_ytb_cluster}
\end{figure}

\begin{table}[h!]
    \centering
    \scriptsize
\begin{tabular}{lrrlll}
\toprule
 & \textbf{Count} & \textbf{(\%)} & \textbf{Topic Name} & \textbf{Author-Coded Group} & \textbf{Representation} \\
\midrule
-1 & 4131 & 25.889317 & Outlier & Outlier & - \\
0 & 2319 & 14.525980 & Collaboration 1 & Collaboration & [family, collabs, anny, turtle, drama] \\
1 & 1824 & 11.434746 & Rude & Rude/Mean & [sibling, chaotic, fun, unapologetically, unique] \\
2 & 1497 & 9.378935 & Random & Random/Chaotic/Unpredictable & [randomness, chaos, unlike, random, world] \\
3 & 1323 & 8.298412 & Swearing 1 & Rude/Mean & [safe, swearing, menace, mode, tune] \\
4 & 1086 & 6.815747 & Mischievous & Random/Chaotic/Unpredictable & [twisted, mischievous, tune, love, play] \\
5 & 906 & 5.690125 & Robotic Voice 1 & Voice/Singing & [generated, vocal, sounds, robotic, emotion] \\
6 & 801 & 5.026778 & Robotic Voice 2 & Voice/Singing & [vocal, generated, expressive, amazing, patterns] \\
7 & 339 & 2.136616 & With Filian & Collaboration & [filian, collabs, cheese, interactions, team] \\
8 & 255 & 1.606690 & Swearing 2 & Rude/Mean & [side, bad, swearing, chaotic, edgy] \\
9 & 249 & 1.570986 & Dumb AI 1 & "Dumb AI" & [steel, feathers, kilogram, physics, heavier] \\
10 & 240 & 1.503336 & Dumb AI 2 & "Dumb AI" & [mistakes, quirks, mess, forget, slip] \\
11 & 201 & 1.268439 & Roast & Rude/Mean & [roasts, honest, setups, harsh, racism] \\
12 & 171 & 1.086160 & Dumb Gameplay & "Dumb AI" & [gameplay, games, time, cool, challenges] \\
13 & 159 & 1.007235 & Math & "Dumb AI" & [basic, 19, mistakes, mess, calculations] \\
14 & 156 & 0.994081 & Collaboration 2 & Collaboration & [room, roasting, collabs, field, harsh] \\
15 & 141 & 0.888847 & Food & "Dumb AI" & [cookies, recipes, bake, attempts, tasks] \\
16 & 138 & 0.877572 & Mincraft & "Dumb AI" & [minecraft, games, bag, plays,  bot] \\
\bottomrule
\end{tabular}
    \caption{Topic information of the topic model for RQ1b: How do viewers perceive the inappropriate, irregular, and other AI-driven quirky behaviors (\ie the ``bad'' behaviors) of AI-driven VTubers?  With YouTube Data.}
    \label{tab:rq1b_ytb_table}
\end{table}

\newpage

\begin{figure}[h!]
    \centering
    \includegraphics[width=\linewidth]{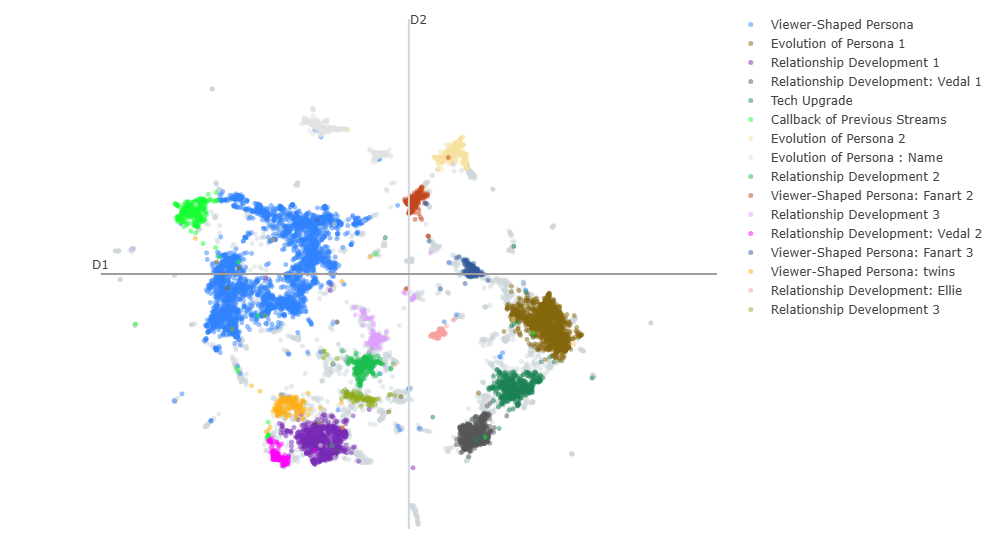}
    \caption{Topic visualization of the topic model for RQ2: How do viewers perceive the consistency and evolution of the virtual personas of AI-driven VTubers? With Reddit Data.}
    \label{fig:rq2_rdt_cluster}
\end{figure}

\begin{table}[h!]
    \centering
    \scriptsize
    \begin{tabular}{lrrlll}
\toprule
 & \textbf{Count} & \textbf{(\%)} & \textbf{Topic Name} & \textbf{Author-Coded Group} & \textbf{Representation} \\
\midrule
-1 & 3371 & 26.323598 & Outlier & Outlier & - \\
0 & 3397 & 26.526628 & Viewer-Shaped Persona & Viewer-Shaped Persona & [fanart, designs, interpretations, cute, love] \\
1 & 1381 & 10.784007 & Evolution of Persona 1 & Evolution of Persona & [self, notice, makes, time,  evolving] \\
2 & 792 & 6.184601 & Relationship Development 1 & Development of Relationships & [father, twins, relationships, neuroverse, interesting] \\
3 & 556 & 4.341715 & Relationship Development: Vedal 1 & Development of Relationships & [creator, bond, parent, shift, feel] \\
4 & 492 & 3.841949 & Tech Upgrade & Changes due to Tech Upgrade & [programming, serious, chat, original, capabilities] \\
5 & 467 & 3.646728 & Callback of Previous Streams & Callback of Content in Previous Streams & [sister, birthday, song, villain, vibe] \\
6 & 429 & 3.349992 & Evolution of Persona 2 & Evolution of Persona & [schoolgirl, original, self,  deviate, acting] \\
7 & 382 & 2.982977 & Evolution of Persona : Name & Evolution of Persona & [name, evilyn, samantha, kayori, evie] \\
8 & 324 & 2.530064 & Relationship Development 2 & Development of Relationships & [family, relationships,  neuroverse, roles, lore] \\
9 & 245 & 1.913166 & Viewer-Shaped Persona: Fanart 1 & Viewer-Shaped Persona & [fanart, created, version, twist, reinterpretations] \\
10 & 199 & 1.553959 & Relationship Development 3 & Development of Relationships & [family, role, neuroverse, relationship, sweet] \\
11 & 188 & 1.468062 & Relationship Development: Vedal 2 & Development of Relationships & [birthday, something, daddy,  bald, cold] \\
12 & 181 & 1.413400 & Viewer-Shaped Persona: Fanart 2 & Viewer-Shaped Persona & [look, fanart, creative, blue, versions] \\
13 & 176 & 1.374356 & Viewer-Shaped Persona: twins & Viewer-Shaped Persona & [roles, twin, beyond, underdog, unique] \\
14 & 117 & 0.913634 & Relationship Development: Ellie & Development of Relationships & [ellie, robot, expertise, relationships, collabs] \\
15 & 109 & 0.851164 & Relationship Development 3 & Development of Relationships & [relationships, collab, neuroverse, interactions] \\
\bottomrule
\end{tabular}
    \caption{Topic information of the topic model for RQ2: How do viewers perceive the consistency and evolution of the virtual personas of AI-driven VTubers? With Reddit Data.}
    \label{tab:rq2_rdt_table}
\end{table}

\newpage

\begin{figure}[h!]
    \centering
    \includegraphics[width=\linewidth]{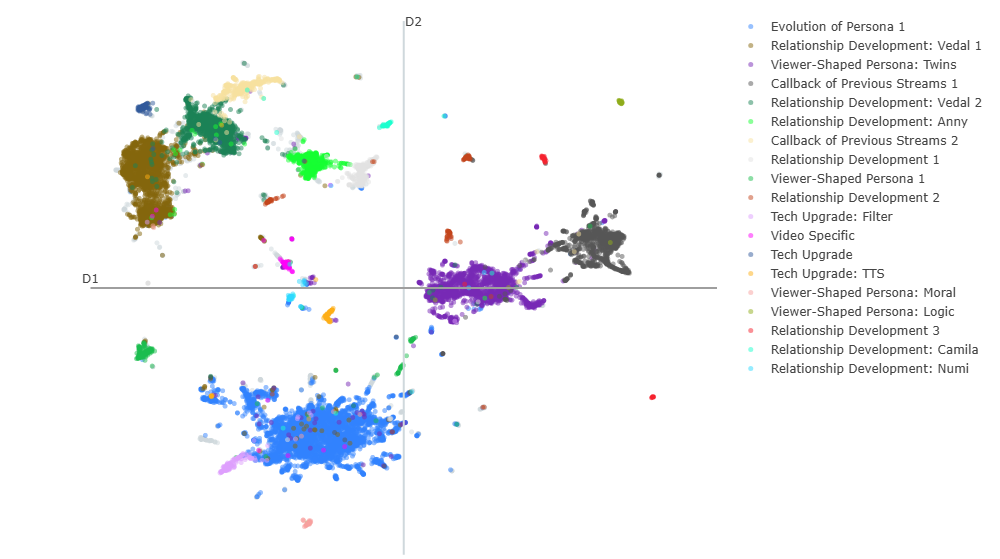}
    \caption{Topic visualization of the topic model for RQ2: How do viewers perceive the consistency and evolution of the virtual personas of AI-driven VTubers? With YouTube Data.}
    \label{fig:rq2_ytb_cluster}
\end{figure}

\begin{table}[h!]
    \centering
    \scriptsize
    \begin{tabular}{lrrlll}
\toprule
 & \textbf{Count} & \textbf{(\%)} & \textbf{Topic Name} & \textbf{Author-Coded Group} & \textbf{Representation} \\
\midrule
-1 & 1492 & 7.296557 & Outlier & Outlier & - \\
0 & 5921 & 28.956377 & Evolution of Persona 1 & Evolution of Persona & [original, beyond, breaks, evolving, think] \\
1 & 2939 & 14.373044 & Relationship Development: Vedal 1 & Development of Relationships & [father, relationship, creation, time, feels] \\
2 & 2680 & 13.106416 & Viewer-Shaped Persona: Twins & Viewer-Shaped Persona & [roles, distinct,  twins, act, cool] \\
3 & 1766 & 8.636541 & Callback of Previous Streams 1 & Callback of Content in Previous Streams & [twin, birthday, development, villain, harpoon] \\
4 & 1598 & 7.814945 & Relationship Development: Vedal 2 & Development of Relationships & [father, twins,  creations, kids, hilarious] \\
5 & 775 & 3.790102 & Relationship Development: Anny & Development of Relationships & [anny, relationships, roles, drama, lore] \\
6 & 678 & 3.315728 & Callback of Previous Streams 2 & Callback of Content in Previous Streams & [birthday, pipe, castle, gifts, last] \\
7 & 495 & 2.420775 & Relationship Development 1 & Development of Relationships & [neuroverse, fox, auntie, relationships, designs] \\
8 & 430 & 2.102895 & Viewer-Shaped Persona 1 & Viewer-Shaped Persona & [angel, singing, beyond, flipped, feels] \\
9 & 289 & 1.413341 & Relationship Development 2 & Development of Relationships & [koko, layna,  collabs, relationships, nurturing] \\
10 & 255 & 1.247066 & Tech Upgrade: Filter & Changes due to Tech Upgrade & [filters, bypassing, free, strictly, breaking] \\
11 & 187 & 0.914515 & Video Specific & Misc/Video Specific & [collabs, chess, playful, feel, evolved] \\
12 & 159 & 0.777582 & Tech Upgrade & Changes due to Tech Upgrade & [avatar, programmer, v2, v3, vrchat] \\
13 & 154 & 0.753130 & Tech Upgrade: TTS & Changes due to Tech Upgrade & [tts, vocal, voice, v3, sounds] \\
14 & 139 & 0.679773 & Viewer-Shaped Persona: Moral & Viewer-Shaped Persona & [trolley, test, moral, dilemmas, psychopath] \\
15 & 131 & 0.640649 & Viewer-Shaped Persona: Logic & Viewer-Shaped Persona & [physics, logic, math, facts, smart] \\
16 & 131 & 0.640649 & Relationship Development 3 & Development of Relationships & [cerber, miniko, collab, partner, friends] \\
17 & 119 & 0.581964 & Relationship Development: Camila & Development of Relationships & [camila, chicken, collabs, members, twins] \\
18 & 110 & 0.537950 & Relationship Development: Numi & Development of Relationships & [numi, collabs, pattern, interactions, fun] \\
\bottomrule
\end{tabular}
    \caption{Topic information of the topic model for RQ2: How do viewers perceive the consistency and evolution of the virtual personas of AI-driven VTubers? With YouTube Data.}
    \label{tab:rq2_ytb_table}
\end{table}
\newpage

\begin{figure}[h!]
    \centering
    \includegraphics[width=\linewidth]{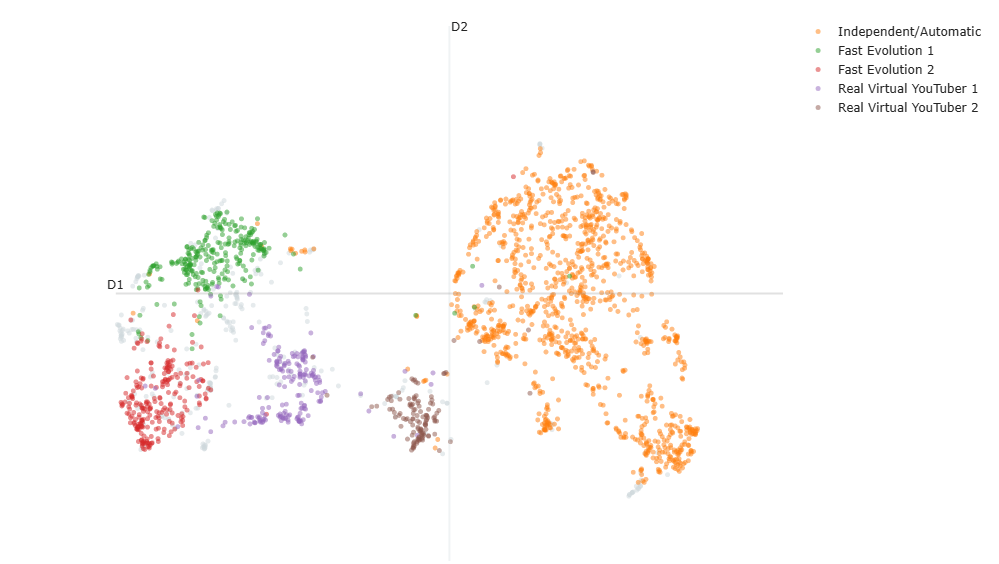}
    \caption{Topic visualization of the topic model for RQ3a: Viewers believe that AI will soon be able to replace the role of Nakanohito in the VTuber industry. With Reddit Data.}
    \label{fig:rq3ay_rdt_cluster}
\end{figure}

\begin{table}[h!]
    \centering
    \scriptsize
    \begin{tabular}{lrrlll}
\toprule
 & \textbf{Count} & \textbf{(\%)} & \textbf{Topic Name} & \textbf{Author-Coded Group} & \textbf{Representation} \\
\midrule
-1 & 276 & 12.354521 & Outlier & Outlier & -  \\
0 & 1127 & 50.447628 & Independent/Automatic & Independent/Automatic & [future, ability, replace, auto, interact] \\
1 & 280 & 12.533572 & Fast Evolution 1 & Fast Evolution & [technology, proves, ai, evolve, v3] \\
2 & 237 & 10.631155 & Fast Evolution 2 & Fast Evolution & [future, advance, evolve, become, ability] \\
3 & 190 & 8.504924 & Real Virtual YouTuber 1 & Real Virtual & [virtual, human, real, true, rely] \\
4 & 123 & 5.528201 & Real Virtual YouTuber 2 & Real Virtual & [virtual, original, literally, ai, anime] \\
\bottomrule
\end{tabular}
    \caption{Topic information of the topic model for RQ3a: Viewers believe that AI will soon be able to replace the role of Nakanohito in the VTuber industry. With Reddit Data.}
    \label{tab:rq3ay_rdt_table}
\end{table}

\newpage

\begin{figure}[h!]
    \centering
    \includegraphics[width=\linewidth]{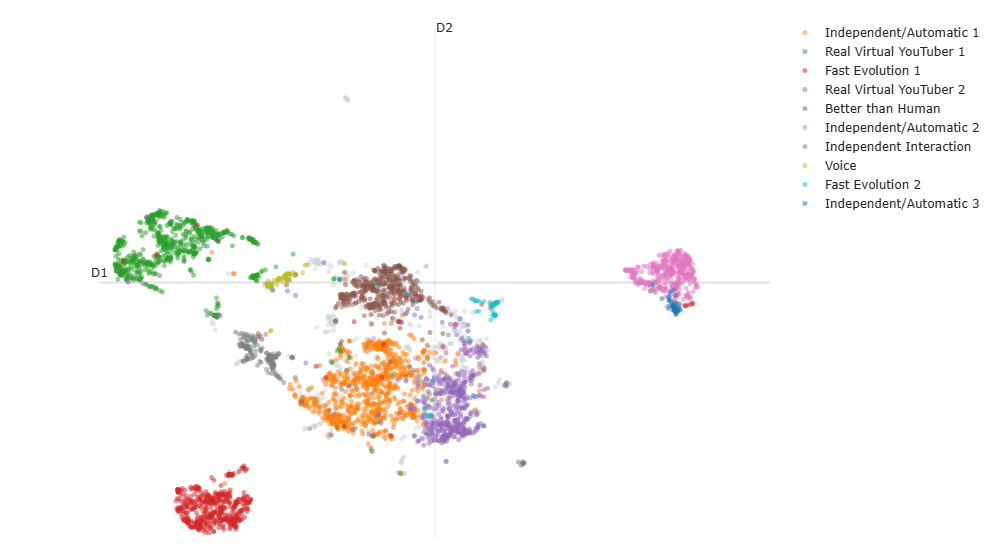}
    \caption{Topic visualization of the topic model for RQ3a: Viewers believe that AI will soon be able to replace the role of Nakanohito in the VTuber industry. With YouTube Data.}
    \label{fig:rq3ay_ytb_cluster}
\end{figure}

\begin{table}[h!]
    \centering
    \scriptsize
    \begin{tabular}{lrrlll}
\toprule
 & \textbf{Count} & \textbf{(\%)} & \textbf{Topic Name} & \textbf{Author-Coded Group} & \textbf{Representation} \\
\midrule
-1 & 480 & 17.102315 & Outlier & Outlier & - \\
0 & 512 & 18.211885 & Independent/Automatic 1 & Independent/Automatic & [ability, need, auto, interactions, operate] \\
1 & 449 & 15.992745 & Real Virtual YouTuber 1 & Real Virtual & [love, humans, virtual, inside, real] \\
2 & 312 & 11.106369 & Fast Evolution 1 & Fast Evolution & [evolution, technology, represents, ability, advanced] \\
3 & 307 & 10.935666 & Real Virtual YouTuber 2 & Real Virtual & [nature,  proves, self, virtual, fact] \\
4 & 280 & 9.975461 & Better than Human & Real Virtual & [engaging, interactions, entertaining, interact, future] \\
5 & 262 & 9.345994 & Independent/Automatic 2 & Independent/Automatic & [creator, independently, operate,  self,  auto] \\
6 & 93 & 3.318041 & Independent Interaction & Independent/Automatic & [twitch, ability, chat, check, interact] \\
7 & 45 & 1.611010 & Voice & Real Virtual & [voice, singing, sing, samples, vocal] \\
8 & 35 & 1.248266 & Fast Evolution 2 & Fast Evolution & [suggesting, reinforces, supporting, program, ai] \\
9 & 32 & 1.152246 & Independent/Automatic 3 & Independent/Automatic & [assistants, work, able, operation, solo] \\
\bottomrule
\end{tabular}
    \caption{Topic information of the topic model for RQ3a: Viewers believe that AI will soon be able to replace the role of Nakanohito in the VTuber industry. With YouTube Data.}
    \label{tab:rq3ay_ytb_table}
\end{table}

\newpage

\begin{figure}[h!]
    \centering
    \includegraphics[width=\linewidth]{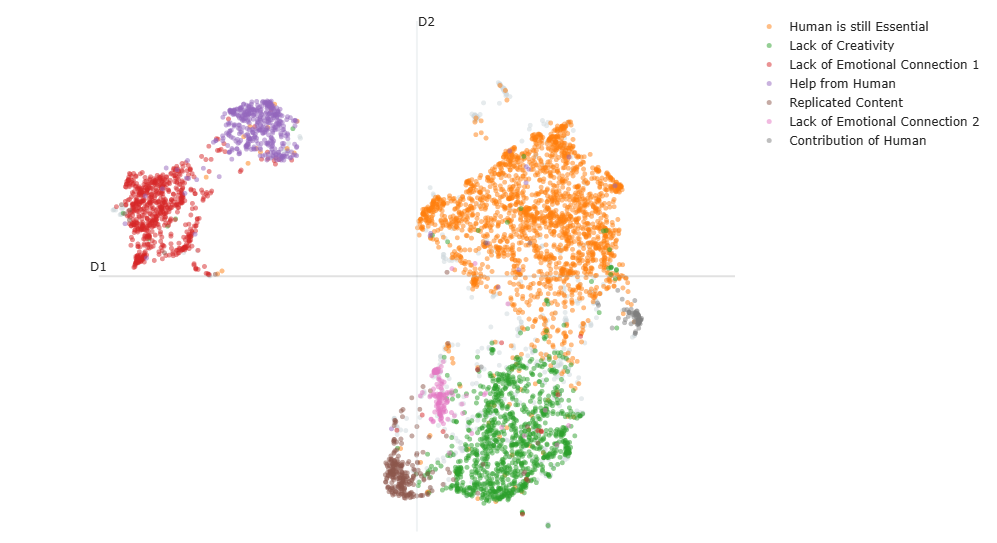}
    \caption{Topic visualization of the topic model for RQ3a: Viewers do not believe that AI will soon be able to replace the role of Nakanohito in the VTuber industry. With Reddit Data.}
    \label{fig:rq3an_rdt_cluster}
\end{figure}

\begin{table}[h!]
    \centering
    \scriptsize
    \begin{tabular}{lrrlll}
\toprule
 & \textbf{Count} & \textbf{(\%)} & \textbf{Topic Name} & \textbf{Author-Coded Group} & \textbf{Representation} \\
\midrule
-1 & 533 & 10.188283 & Outlier & Outlier & - \\
0 & 2007 & 38.363758 & Human is still Essential & Human in the Loop & [essential, fact, element, creators, involvement] \\
1 & 1079 & 20.625060 & Lack of Creativity & Lack of Diversity/Depth & [replicate, depth, authenticity, unique, creativity] \\
2 & 705 & 13.485616 & Lack of Emotional Connection 1 & Lack of Emotional Connection & [emotional, connection, emotion, feels, avatar] \\
3 & 442 & 8.448820 & Help from Human & Human in the Loop & [existence, essential, success, maintainer, community] \\
4 & 259 & 4.950779 & Replicated Content & Lack of Diversity/Depth & [repeat, unique, replace, algorithms, matter] \\
5 & 129 & 2.475389 & Lack of Emotional Connection 2 & Lack of Emotional Connection & [feeling, connection, something, people, special] \\
6 & 76 & 1.462296 & Contribution of Human & Human in the Loop & [anny, mother, contributions, vital, support] \\
\bottomrule
\end{tabular}
    \caption{Topic information of the topic model for RQ3a: Viewers do not believe that AI will soon be able to replace the role of Nakanohito in the VTuber industry. With Reddit Data.}
    \label{tab:rq3an_rdt_table}
\end{table}

\newpage

\begin{figure}[h!]
    \centering
    \includegraphics[width=\linewidth]{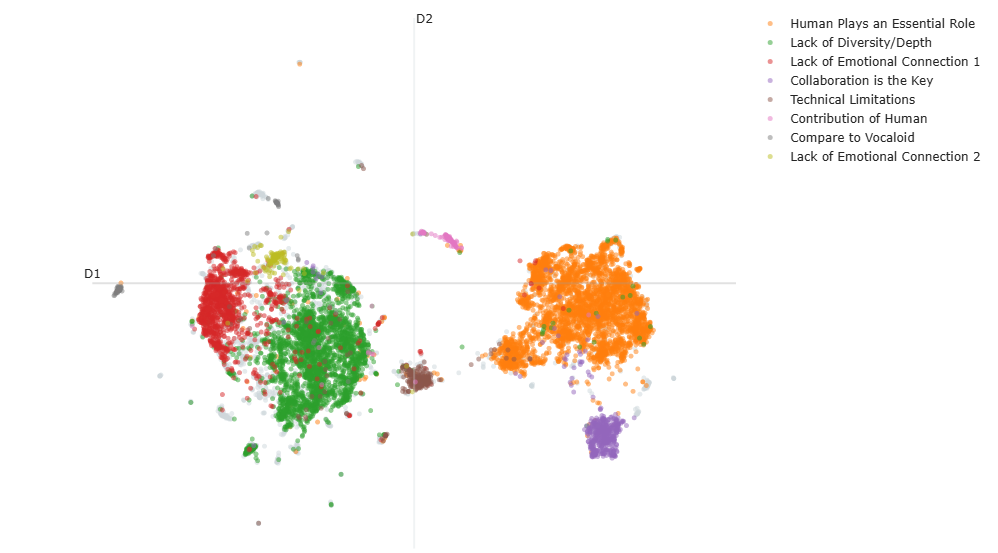}
    \caption{Topic visualization of the topic model for RQ3a: Viewers do not believe that AI will soon be able to replace the role of Nakanohito in the VTuber industry. With YouTube Data.}
    \label{fig:rq3an_ytb_cluster}
\end{figure}

\begin{table}[h!]
    \centering
    \scriptsize
    \begin{tabular}{lrrlll}
\toprule
 & \textbf{Count} & \textbf{(\%)} & \textbf{Topic Name} & \textbf{Author-Coded Group} & \textbf{Representation} \\
\midrule
-1 & 1117 & 16.624281 & Outlier & Outlier & - \\
0 & 2272 & 33.801811 & Human Plays an Essential Role & Human in the Loop & [essential, rely, element, existence,  function] \\
1 & 1751 & 26.047383 & Lack of Diversity/Depth & Lack of Diversity/Depth & [real, depth, replicated, replace, lacks] \\
2 & 755 & 11.239013 & Lack of Emotional Connection 1 & Lack of Emotional Connection & [connection, authenticity, unique, touch, soul] \\
3 & 350 & 5.206800 & Collaboration is the Key & Human in the Loop & [creators, essential, neuroverse, involvement, work] \\
4 & 203 & 3.029492 & Technical Limitations & Lack of Diversity/Depth & [challenges, oversight, behavior, flaw, chat] \\
5 & 100 & 1.494668 & Contribution of Human & Human in the Loop & [mom, contribute, creation, neuroverse, anny] \\
6 & 95 & 1.418819 & Compare to Vocaloid & Misc/Video Specific & [miku, legacy, cultural, hatsune, vocaloids] \\
7 & 76 & 1.137733 & Lack of Emotional Connection 2 & Lack of Emotional Connection & [future, value, connection, improvement, touch] \\
\bottomrule
\end{tabular}
    \caption{Topic information of the topic model for RQ3a: Viewers do not believe that AI will soon be able to replace the role of Nakanohito in the VTuber industry. With YouTube Data.}
    \label{tab:rq3an_ytb_table}
\end{table}

\newpage

\begin{figure}[h!]
    \centering
    \includegraphics[width=\linewidth]{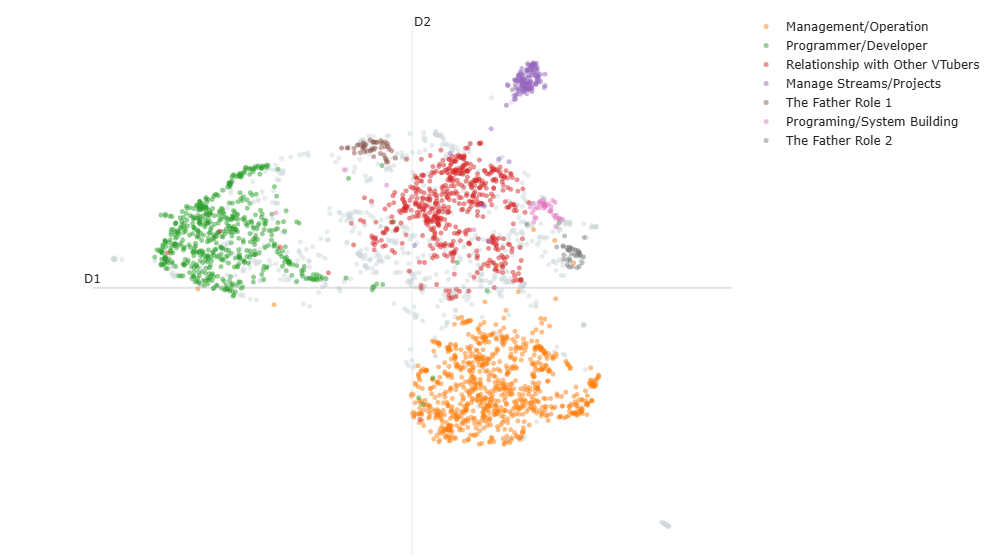}
    \caption{Topic visualization of the topic model for RQ3b:  What concerns do viewers have regarding the creator/maintainer/operator of AI-driven VTubers? With Reddit Data.}
    \label{fig:rq3b_rdt_cluster}
\end{figure}

\begin{table}[h!]
    \centering
    \scriptsize
    \begin{tabular}{lrrlll}
\toprule
 & \textbf{Count} & \textbf{(\%)} & \textbf{Topic Name} & \textbf{Author-Coded Group} & \textbf{Representation} \\
\midrule
-1 & 731 & 25.025676 & Outlier & Outlier & - \\
0 & 813 & 27.832934 & Management/Operation & Key Decision Maker & [control, managing, decision, operations, direction] \\
1 & 578 & 19.787744 & Technical Control over AI Persona & Technical Control over AI Persona & [control, creator, concept, actions, influence] \\
2 & 529 & 18.110236 & Relationship with Other VTubers & Relationship with Other VTubers & [events, collabs, neuroverse, projects, ships] \\
3 & 140 & 4.792879 & Manage Streams/Projects & Key Decision Maker & [operator, direction, behavior, managing, channels] \\
4 & 45 & 1.540568 & The Father Role 1 & The Father Role & [father, parent,  daughters, dad, child] \\
5 & 43 & 1.472099 & Program/System & Technical Control over AI Persona & [real, systems, consciousness, functionality, programming] \\
6 & 42 & 1.437864 & The Father Role 2 & The Father Role & [turtle, daddy,  daughters, dad, child] \\
\bottomrule
\end{tabular}
    \caption{Topic information of the topic model for RQ3b:  What concerns do viewers have regarding the creator/maintainer/operator of AI-driven VTubers? With Reddit Data.}
    \label{tab:rq3b_rdt_table}
\end{table}

\newpage

\begin{figure}[h!]
    \centering
    \includegraphics[width=\linewidth]{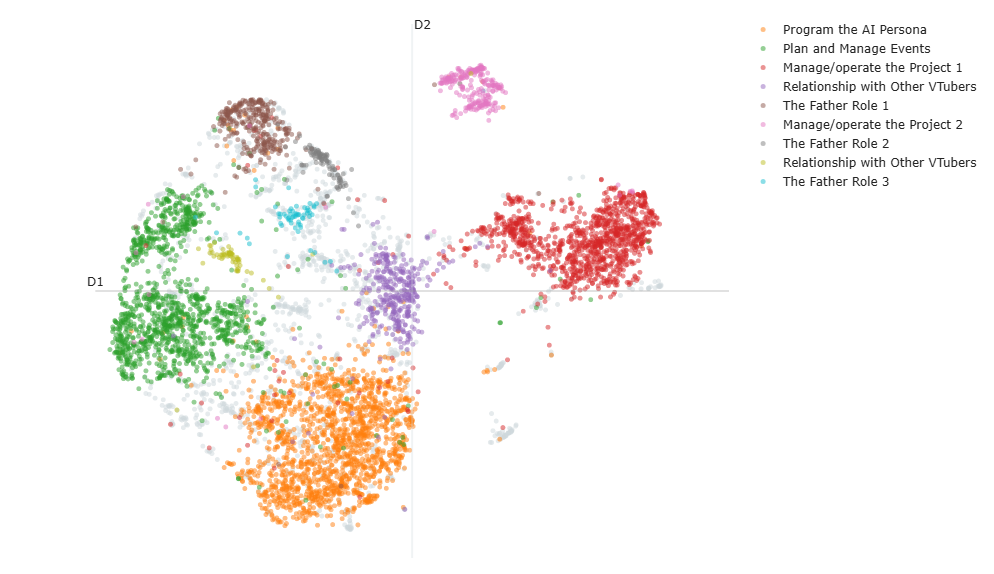}
    \caption{Topic visualization of the topic model for RQ3b:  What concerns do viewers have regarding the creator/maintainer/operator of AI-driven VTubers? With YouTube Data.}
    \label{fig:rq3b_ytb_cluster}
\end{figure}

\begin{table}[h!]
    \centering
    \scriptsize
    \begin{tabular}{lrrlll}
\toprule
 & \textbf{Count} & \textbf{(\%)} & \textbf{Topic Name} & \textbf{Author-Coded Group} & \textbf{Representation} \\
\midrule
-1 & 1226 & 18.910998 & Outlier & Outlier & - \\
0 & 1642 & 25.327780 & Technical Control over AI Persona & Technical Control over AI Persona & [programming, control, actions, behavior, creator] \\
1 & 1277 & 19.697671 & Plan and Manage Events & Key Decision Maker & [events, managing, projects, operate, operations] \\
2 & 1153 & 17.784976 & Manage/operate the Project 1 & Key Decision Maker & [person, projects, operations, events, acting] \\
3 & 372 & 5.738084 & Relationship with Other VTubers & Relationship with Other VTubers & [collaborations, collabs, neuroverse, person, ships] \\
4 & 319 & 4.920561 & The Father Role 1 & The Father Role & [dad, daughter, parent,  creator, concept] \\
5 & 269 & 4.149314 & Manage/operate the Project 2 & Key Decision Maker & [idea, projects, planning, operates, actually] \\
6 & 99 & 1.527071 & The Father Role 2 & The Father Role & [parental, dad, joke, tied, responsibility] \\
7 & 65 & 1.002622 & Relationship with Other VTubers & Relationship with Other VTubers & [anny, camila, ships, miniko, cerber] \\
8 & 61 & 0.940922 & The Father Role 3 & The Father Role & [father, figure,parent, parenthood, children] \\
\bottomrule
\end{tabular}
    \caption{Topic information of the topic model for RQ3b:  What concerns do viewers have regarding the creator/maintainer/operator of AI-driven VTubers? With YouTube Data.}
    \label{tab:rq3b_ytb_table}
\end{table}

\end{document}